\documentclass[]{aa}  

\usepackage{graphicx}
\usepackage{txfonts}
\usepackage{xcolor}
\usepackage{textcomp}
\usepackage{amsmath}	% Advanced maths commands
\usepackage{amssymb}	% Extra maths symbols
\usepackage{upgreek}
\usepackage{placeins}
\usepackage{hyperref}

\newcommand{\orcid}[1]{\href{https://orcid.org/#1}{\includegraphics[width=10pt]{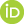}}}

\begin{document} 

   \title{SN 2023gpw: exploring the diversity and power sources of hydrogen-rich superluminous supernovae}

   %\subtitle{}

   \author{T. Kangas
          \inst{1,2}\orcid{0000-0002-5477-0217}
          \and
          P. Charalampopoulos\inst{2}\orcid{0000-0002-0326-6715}
          \and
          T. Nagao\inst{2,3,4}\orcid{0000-0002-3933-7861}
          \and
          L. Yan\inst{5}\orcid{0000-0003-1710-9339}
          \and
          M. Stritzinger\inst{6}\orcid{0000-0002-5571-1833}
          \and
          S. Schulze\inst{7}\orcid{0000-0001-6797-1889}
          \and
          K. Das\inst{8}\orcid{0000-0001-8372-997X}
          \and
          N. Elias-Rosa\inst{9,10}\orcid{0000-0002-1381-9125}
          \and
          C. Fremling\inst{11}\orcid{0000-0002-4223-103X}
          \and
          D. Perley\inst{12}\orcid{0000-0001-8472-1996}
          \and
          J. Sollerman\inst{13}\orcid{0000-0003-1546-6615}
          \and
          T. E. M\"uller-Bravo\inst{14,15}\orcid{0000-0003-3939-7167}
          \and
          L. Galbany\inst{10,16}\orcid{0000-0002-1296-6887}
          \and
          S. L. Groom\inst{17}\orcid{0000-0001-5668-3507}
          \and
          C. P. Guti\'{e}rrez\inst{16,10}\orcid{0000-0003-2375-2064}
          \and
          M. M. Kasliwal\inst{11}\orcid{0000-0002-5619-4938}
          \and
          R. Kotak\inst{2}\orcid{0000-0001-5455-3653}
          \and
          R. R. Laher\inst{17}\orcid{0000-0003-2451-5482}
          \and
          P. Lundqvist\inst{13}\orcid{0000-0002-3664-8082}
          \and
          S. Mattila\inst{2}\orcid{0000-0001-7497-2994}
          \and
          R. Smith\inst{5}\orcid{0000-0001-7062-9726}
          }

   \institute{Finnish Centre for Astronomy with ESO (FINCA), FI-20014 University of Turku, Finland
            \email{tjakangas@gmail.com}
            \and
            Department of Physics and Astronomy, FI-20014 University of Turku, Finland 
            \and
            Aalto University Metsähovi Radio Observatory, Metsähovintie 114, 02540 Kylmälä, Finland
            \and
            Aalto University Department of Electronics and Nanoengineering, PO Box 15500, 00076 Aalto, Finland
            \and
            The Caltech Optical Observatories, California Institute of Technology, Pasadena, CA 91125, USA
            \and
            Department of Physics and Astronomy, Aarhus University, Ny Munkegade 120, 8000 Aarhus C, Denmark
            \and
            Center for Interdisciplinary Exploration and Research in Astrophysics (CIERA), Northwestern University, 1800 Sherman Ave, Evanston, IL 60201, USA
            \and
            Cahill Center for Astrophysics, California Institute of Technology, MC 249-17, 1200 East California Boulevard, Pasadena, CA 91125, USA
            \and
            INAF - Osservatorio Astronomico di Padova, Vicolo dell’Osservatorio 5, I-35122 Padova, Italy
            \and
            Institute of Space Sciences (ICE, CSIC), Campus UAB, Carrer de Can Magrans s/n, 08193 Barcelona, Spain
            \and
            Division of Physics, Mathematics, and Astronomy, California Institute of Technology, Pasadena, CA 91125, USA
            \and
            Astrophysics Research Institute, Liverpool John Moores University, Liverpool Science Park, 146 Brownlow Hill, Liverpool L3 5RF, UK
            \and
            Department of Astronomy, The Oskar Klein Centre, Stockholm University, AlbaNova, SE-106 91 Stockholm, Sweden
            \and
            School of Physics, Trinity College Dublin, The University of Dublin, Dublin 2, Ireland
            \and
            Instituto de Ciencias Exactas y Naturales (ICEN), Universidad Arturo Prat, Chile
            \and
            Institut d’Estudis Espacials de Catalunya (IEEC), Edifici RDIT, Campus UPC, 08860 Castelldefels (Barcelona), Spain
            \and
            IPAC, California Institute of Technology, 1200 E. California Blvd, Pasadena, CA 91125, USA
            \and
            Department of Physics, University of Warwick, Gibbet Hill Road, Coventry CV4 7AL, UK
            }

   \date{Received ,; accepted ,}
 
  \abstract
   {
    We present our observations and analysis of SN~2023gpw, a hydrogen-rich superluminous supernova (SLSN~II) with broad emission lines in its post-peak spectra. Unlike previously observed SLSNe~II, its light curve suggests an abrupt drop during a solar conjunction between $\sim$80 and $\sim$180~d after the light-curve peak, possibly analogous to a normal hydrogen-rich supernova (SN). Spectra taken at and before the peak show hydrogen and helium `flash' emission lines attributed to early interaction with a dense confined circumstellar medium (CSM). A well-observed ultraviolet excess appears as these lines disappear, also as a result of CSM interaction. The blackbody photosphere expands roughly at the same velocity throughout the observations, indicating little or no bulk deceleration. This velocity is much higher than what is seen in spectral lines, suggesting asymmetry in the ejecta. The high total radiated energy ($\gtrsim9\times10^{50}$~erg) and aforementioned lack of bulk deceleration in SN~2023gpw are difficult to reconcile with a neutrino-driven SN simply combined with efficient conversion from kinetic energy to emission through interaction. This suggests an additional energy source such as a central engine. While magnetar-powered models qualitatively similar to SN~2023gpw exist, more modeling work is required to determine if they can reproduce the observed properties in combination with early interaction. The required energy might alternatively be provided by accretion onto a black hole created in the collapse of a massive progenitor star.
   }


   \maketitle
%
%-------------------------------------------------------------------

\section{Introduction}
\label{sec:intro}

Massive stars ($\gtrsim8~\mathrm{M}_\odot$) end their lives in luminous explosions known as supernovae (SNe). These explosions are divided into two main classes: hydrogen-poor SNe~I and hydrogen-rich SNe~II. For both types, there is also an analogous type of superluminous supernovae \citep[SLSNe; for reviews see][]{galyam12,galyam19,moriya24}, which reach peak absolute magnitudes of $\lesssim-20$~mag, a few or even several magnitudes more luminous than their normal counterparts -- though we note that for H-rich SLSNe this limit is somewhat arbitrary and some overlap is expected. Occasionally the peak magnitude can even approach $\sim-23$~mag \citep[e.g.][]{kangas22}. Such luminosities, often paired with a relatively slow evolution, result in enormous energy requirements that the normal SN powering mechanisms -- the decay of $^{56}$Ni and the cooling of the shock-heated envelope -- cannot attain, and other power sources are needed. 

Such a SN could in principle be a so-called pair-instability SN \citep[PISN; ][]{barkat67,hw02}, where extreme temperatures resulting in pair production are reached at the core, decreasing radiation pressure and inducing a collapse. A PISN could power the luminosity with an extreme $^{56}$Ni mass, but there are only a few candidates for this mechanism \citep[see, e.g.,][]{lunnan16,schulze24}, and even then PISN models do not reproduce all main observables. In most cases, one of two main power sources is considered; either the birth of a strongly-magnetized, fast-rotating neutron star, a millisecond magnetar, which subsequently acts as a central engine by shedding its rotational energy through magnetic braking \citep{kb10,woosley10}; or interaction between the ejecta and a dense circumstellar medium (CSM) resulting in efficient conversion of the kinetic energy of the ejecta into radiation \citep[e.g.][]{cf94,ofek07,sorokina16}. Such interaction is common even among normal SNe~II, where the ejecta may interact with as much as $\sim$0.5~M$_\odot$ of CSM at early times \citep{morozova18}. In addition to the magnetar spin-down, another possible central engine is fallback accretion onto a nascent black hole \citep{dexterkasen13}.

In SLSNe~I, the more extensively studied SLSN type \citep[e.g.][]{quimby11,chen23a}, the dominant mechanism is generally considered to be the magnetar central engine \citep[e.g.][]{inserra13,nicholl15,gomez24}, though circumstellar interaction (CSI) seems to be required to explain some objects \citep[e.g.][]{lunnan18,chen23b}. In particular, undulations commonly seen in SLSN~I light curves \citep[e.g.][]{inserra17,hosseinzadeh22,west23} may require CSI, though variations in the output of the central engine are plausible in some cases as well \citep{chugaiutrobin22,moriya22,chen23b}. 

More luminous analogs to SNe~IIn (i.e., with narrow Balmer lines indicative of CSI) form the majority of hydrogen-rich SLSNe \citep{galyam19,kangas22,pessi24}. As shown by \citet{hiramatsu24}, the distribution of SNe IIn and their superluminous equivalents is continuous with no clear boundary. However, some SLSNe instead exhibit broad Balmer lines; the prototype of this subtype is SN~2008es \citep{gezari09,miller09}. In these objects, the lack of narrow lines from dense, slow-moving CSM makes the dominant power source somewhat ambiguous, and both magnetars and CSI have been suggested \citep{inserra18,kornpob19}. In this paper, we follow the nomenclature of \citet{galyam19} and \citet{kangas22}: we refer to the broad-lined subtype as SLSNe~II and the narrow-lined one as SLSNe~IIn for brevity.

In the largest sample of SLSNe~II (i.e., not IIn) so far, \citet{kangas22} found sub-groups that spectroscopically resemble either SN~2008es or normal SNe~II; the former group tends to be more luminous at peak. While signs of CSI, such as an excess of ultraviolet (UV) emission over a blackbody, were seen, magnetar and CSI models tended to be equally effective in reproducing the light curves. The most luminous SLSN~II was estimated to emit $\gtrsim3 \times 10^{51}$~erg. This poses a problem for the normal, neutrino-driven core-collapse mechanism with CSI, as even with full conversion of kinetic energy into radiation, it seems impossible to obtain such an energy \citep{janka12}, and it is possible that some SLSNe~II require both a central engine and CSI. Whichever mechanism is dominant among SLSNe, they tend to occur in faint, low-metallicity host galaxies -- and SLSNe~I and 2008es-like SNe seem to require an even lower metallicity than other SLSNe~II or SLSNe~IIn \citep[e.g.][]{angus16,perley16,schulze21,kangas22}.

In this paper, we present UV, optical and infrared photometric and spectroscopic observations of a SLSN~II, SN~2023gpw. This SN exhibits a clear UV excess and early `flash' emission features similar to those seen early on in normal SNe~II, which indicate CSI \citep[e.g.][]{bruch21,wynn24}. We analyze the light curve and spectral evolution, comparing them to other SLSNe~II, in order to assess the physical nature of the transient. We present the observations we use, including proprietary and public ones, in Sect. \ref{sec:data}, our analysis in Sect. \ref{sec:analysis}, models of the photometric data of the SN and its host galaxy in Sect. \ref{sec:models}, and a discussion of our results in Sect. \ref{sec:disco}. Finally, our conclusions are presented in Sect. \ref{sec:concl}. Throughout this paper, magnitudes are in the AB system \citep{okegunn83} unless otherwise noted, and we assume a $\Lambda$CDM cosmology with parameters $H_0 = 69.6$~km~s$^{-1}$~Mpc$^{-1}$, $\Omega_M = 0.286$ and $\Omega_\Lambda = 0.714$ \citep{bennett14}. Reported uncertainties correspond to 68\% ($1\sigma$) and upper limits to $3\sigma$. 

%--------------------------------------------------------------------
\section{Observations}
\label{sec:data}

\begin{figure}
\centering
\includegraphics[width=0.7\linewidth]{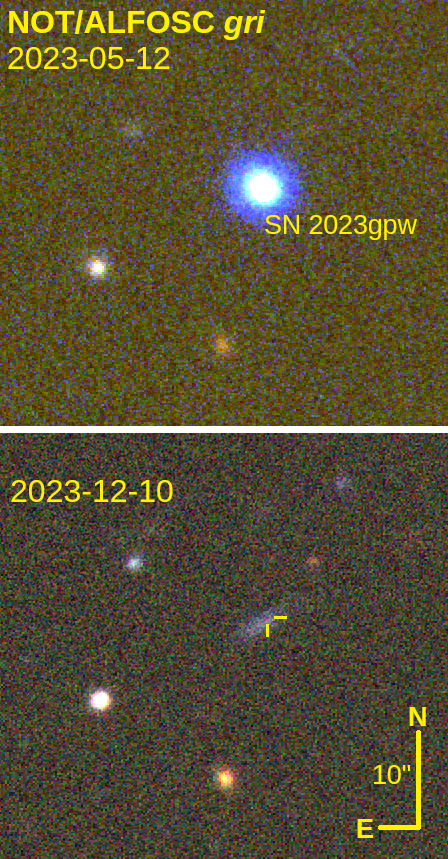}
\caption{Three-color NOT/ALFOSC images (blue: $g$; green: $r$; red: $i$) of SN~2023gpw a week before the light-curve peak (left) and its faint host galaxy, visible  at 189~d after the peak and with little contribution from the SN (right). The position of the SN in the galaxy is indicated, and the vertical bar corresponds to 10 arcseconds.}
\label{fig:snfield}
\end{figure}

The discovery of SN~2023gpw in the Zwicky Transient Facility \citep[ZTF;][]{bellm19,graham19} on MJD\,=\,60051.3 was reported by \citet{alercereport}. It was found at RA = 13:02:18.70, Dec = $-$05:51:10.95 using the Automatic Learning for the Rapid Classification of Events (ALeRCE) broker \citep{forster21}\footnote{\url{https://alerce.science/}} and given the internal designation ZTF23aagklwb. We show the vicinity and faint host galaxy of the SN in Fig. \ref{fig:snfield}.  The photometric properties and the projected location near the nucleus of its host galaxy made the transient a candidate for a tidal disruption event (TDE) and prompted a spectroscopic observation as part of a TDE-related program (proposal P67-021, PI Charalampopoulos). The spectrum was taken on MJD\,=\,60071.1 using the Alhambra Faint Object Spectrograph and Camera (ALFOSC) on the 2.56-m Nordic Optical Telescope (NOT) located at the Roque de los Muchachos Observatory, La Palma. The spectrum initially showed similarity to either a TDE or a SLSN~II; the public classification of the event as a SLSN~II and the SN~2023gpw designation \citep{kravtsov23} was done later, on MJD\,=\,60101.2, as part of the Public ESO Spectroscopic Survey for Transient Objects \citep[PESSTO;][]{pessto}.\footnote{\url{https://www.pessto.org/}} Based on the initial spectrum, we triggered a multi-wavelength follow-up as described below. The reduction of the resulting data is described in Appendix \ref{sec:redu}.

Near-ultraviolet (NUV) observations of SN 2023gpw were performed with the Neil Gehrels \emph{Swift} Observatory in the $W1$, $M2$ and $W2$ filters between MJD\,=\,60075.8 and MJD\,=\,60141.3. A late-time NUV template set containing only the host galaxy was also obtained on MJD\,=\,60681.2. The data were downloaded using the HEASARC Archive.\footnote{\url{https://heasarc.gsfc.nasa.gov/cgi-bin/W3Browse/swift.pl}} Optical imaging of SN~2023gpw was done as part of ZTF using the ZTF Camera \citep{dekany20} on the 48-inch Samuel Oschin Telescope at Palomar; the Spectral Energy Distribution Machine \citep[SEDM;][]{blagorodnova2018} on the Palomar 60\,inch (P60) telescope in the $gri$ bands; and the IO:O instrument on the 2-m robotic Liverpool Telescope \citep[LT;][]{Steele04}, located on La Palma, in the $griz$ bands. Outside ZTF, we used the Las Cumbres Observatory Global Telescope \citep[LCOGT;][]{brown13} network (proposal OPTICON 22A/012, PI Stritzinger) in the $ugri$ bands. We also obtained one epoch of $ugri$ photometry and four epochs of late-time (after MJD\,=\,60289) $gri$ photometry using NOT/ALFOSC as part of the NOT Un-biased Transient Survey 2 (NUTS2) collaboration\footnote{\url{https://nuts.sn.ie/}} (proposal 66-506, PIs Kankare, Stritzinger, Lundqvist). In the near-infrared (NIR), we obtained one epoch in the $JHKs$ bands using the Nordic Optical Telescope near-infrared Camera and spectrograph (NOTCam) on the NOT as part of NUTS2. All our photometric measurements, without $K$-corrections or extinction corrections, are shown in Appendix \ref{sec:tables} in Tables \ref{tab:notcam} to \ref{tab:uvot}. Finally, we acquired one epoch (MJD 60081.01) of $V$ and $R$ band imaging polarimetry with NOT/ALFOSC. All observations were obtained at four half-wave plate angles (0$^{\circ}$, 22.5$^{\circ}$, 45$^{\circ}$, 67.5$^{\circ}$) with an exposure time of 100 seconds per half-wave plate. 

Spectroscopic observations of SN~2023gpw were performed using the following instruments:
\begin{itemize}
    \item three spectra using NOT/ALFOSC: one on MJD\,=\,60071.1 and two as part of NUTS2 on MJD\,=\,60077.0 and 60084.9;
    \item two spectra using the Intermediate Dispersion Spectrograph (IDS) on the 2.5-m Isaac Newton Telescope (INT) on La Palma on MJD\,=\,60097.0 and 60097.9 (proposal C54, PI M\"{u}ller-Bravo);
    \item two spectra using the Double Beam Spectrograph \citep[DBSP;][]{okegunn82} on the Palomar 200-inch Telescope (P200) in Palomar, California, USA, as part of ZTF on MJD\,=\,60116.2 and 60137.3;
    \item 12 spectra using SEDM as part of ZTF between MJD\,=\,60071.2 and 60121.2;
    \item a late-time spectrum using the Optical System for Imaging and low-Intermediate-Resolution Integrated Spectroscopy (OSIRIS+) instrument on the 10.4-m Gran Telescopio Canarias (GTC) on La Palma on MJD\,=\,60386.2 (proposal GTCMULTIPLE2G-24A, PI Elias-Rosa).
\end{itemize}
The properties of these spectra are listed in Table \ref{tab:specs}. 

In addition to proprietary data, we retrieved the public PESSTO classification spectrum of SN~2023gpw \citep{kravtsov23} from the Transient Name Server (TNS).\footnote{\url{https://www.wis-tns.org/}} This spectrum was taken on MJD\,=\,60101.2 with the ESO Faint Object Spectrograph and Camera 2 (EFOSC2) instrument on the New Technology Telescope located in La Silla, Chile. We also obtained the Asteroid Terrestrial-impact Last Alert System \citep[ATLAS;][]{tonry11,smith20} forced photometry \citep{shingles21} of SN~2023gpw, which extends from MJD\,=\,60100.8 to MJD\,=\,60169.0 in the ATLAS wide-band $c$ and $o$ filters. 

%--------------------------------------------------------------------
\section{Analysis}
\label{sec:analysis}

%--------------------------------------------------------------------
\subsection{Light curve}
\label{sec:LC}

\begin{figure*}
\centering
\includegraphics[width=0.9\linewidth]{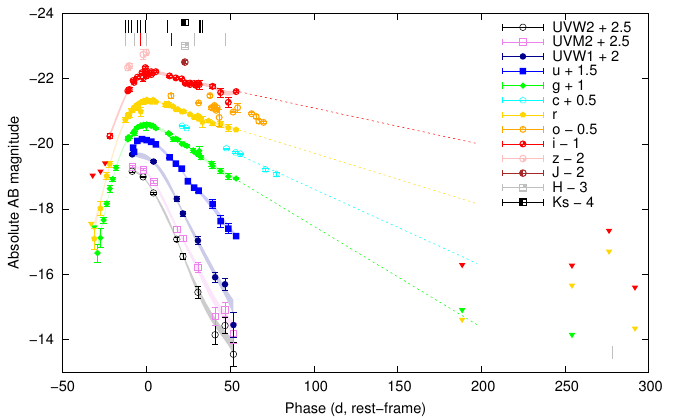}
\caption{Light curve of SN 2023gpw. The zero epoch corresponds to the $g$-band peak on MJD\,=\,60085.0. Triangles denote $3\sigma$ upper limits, and shaded regions denote $1\sigma$ uncertainties of our GP fits. Epochs of spectra are marked with black (SEDM), gray (our other spectra) or blue \citep{kravtsov23} lines, and the epoch of polarimetry is marked with a red line. Dashed lines show the evolution if the pre-gap decline rates had stayed constant during the gap, showing that the decline must have steepened some time after the last detection in the $r$ and $i$ bands.}
\label{fig:absLC}
\end{figure*}

We applied a Milky Way extinction correction of $A_{V,\mathrm{MW}} = 0.097$~mag \citep{schlafly11} to our photometry, using the \citet{gordon23} extinction law (with $R_V = 3.1$). The observed early blue color (see below and Sect. \ref{sec:BB}) and a lack of narrow interstellar absorption lines, especially from Na~{\sc i} (Sect. \ref{sec:specs}), suggest a minimal host-galaxy extinction, and we only correct our photometry for the Milky Way extinction in this paper. Our host-galaxy fitting (Sect. \ref{sec:host}) agrees with this result. 

Due to the heliocentric redshift $z=0.0827\pm0.0002$ of SN~2023gpw based on host-galaxy lines (see Sect. \ref{sec:specs}), we also apply an approximate $K$-correction \citep{hogg02} of $-2.5 \mathrm{log}(1+z) \approx -0.09$~mag (i.e., we add 0.09 to the observed magnitude). This approximation was shown to work reasonably well for SLSNe~II in the optical by \citet{kangas22}. For SN~2023gpw, we used the \texttt{SNAKE} code \citep{inserra18} to determine spectroscopic $K$-corrections in the $g$ and $r$ bands, which we use for light-curve comparisons. These were found to be close to the above approximation ($-0.08$~mag in $g$ and $-0.12$~mag in $r$) around the peak. In the NUV and NIR bands, we do not have spectra available, and in the NIR, only one epoch of photometry for determining the $K$-correction. Therefore, we only apply the approximate correction in all bands, which keeps the object comparable to the literature sample.

Based on the redshift -- in the Galactic standard of rest, $z_\mathrm{GSR} = 0.0824 \pm 0.0002$ --  and the assumed cosmology, we obtain a luminosity distance of $D_L = 377.3\pm1.0$~Mpc and a distance modulus of $\mu = 37.88\pm0.01$~mag. The rest-frame absolute-magnitude light curve is shown in Fig. \ref{fig:absLC}. For clarity, ATLAS magnitudes were averaged when multiple observations were performed within 0.1~d.

We performed numerical interpolation of the light curves using a Gaussian process (GP) regression algorithm \citep{rasmussen06} to estimate the peak magnitudes and epochs. We used the Python-based {\tt george} package \citep{hodlr}, which implements various kernel functions. The widely-used Mat{\'e}rn-3/2 kernel was applied, with a 0.05~mag error floor added to measured magnitudes to avoid overfitting because of individual points, and hyperparameters were not fixed. The $1\sigma$ lower and upper bounds of the peak epoch were conservatively estimated as the epochs when the maximum-likelihood GP fit becomes, respectively, brighter and fainter than the $1\sigma$ lower bound on the peak brightness.

The peak epoch in the $g$ band is MJD$_{g,\mathrm{peak}} = 60085.0^{+3.1}_{-2.9}$, which we use as the reference epoch in the rest of the paper. The peak absolute magnitude in $g$ is $M_{g,\mathrm{peak}} = -21.58\pm0.03$ mag, which makes SN~2023gpw relatively luminous among the SLSNe~II of \citet{kangas22}, roughly on the borderline between SLSNe~II that spectroscopically resemble normal SNe~II and those that resemble SN~2008es. The measured peak epochs and magnitudes in all bands are listed in Table \ref{tab:peaks}. The light curve peaks slightly earlier and is brighter at shorter wavelengths. In the NUV bands the peak epoch cannot be determined due to a lack of pre-peak observations; however, by eye (Fig. \ref{fig:absLC}), the peak seems to be close to the first observed epoch (MJD\,=\,60075.8) in all NUV bands, and certainly earlier than in the optical.

\begin{table}
\centering
\caption{Epochs and absolute magnitudes at peak in the optical bands.}
\label{tab:peaks}
\begin{tabular}{ccc}
\hline
Filter & MJD$_\mathrm{peak}$ & $M_{\mathrm{peak}}$ \\
 &  & (mag) \\
\hline
\hline
$u$ & $60083.9^{+3.3}_{-2.6}$ & $-21.63\pm0.04$ \\
$g$ & $60085.0^{+3.1}_{-2.9}$ & $-21.58\pm0.03$ \\
$r$ & $60086.8^{+4.5}_{-3.4}$ & $-21.32\pm0.03$ \\
$i$ & $60091.5^{+5.3}_{-9.9}$ & $-21.18\pm0.04$ \\
\hline
\end{tabular}
\tablefoot{The $g$-band peak is used as the reference epoch throughout the paper.}
\end{table}

\begin{figure*}
\centering
\includegraphics[width=0.9\linewidth]{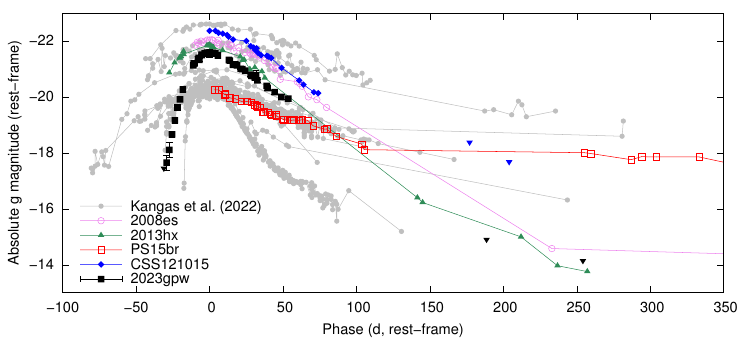} \\
\includegraphics[width=0.9\linewidth]{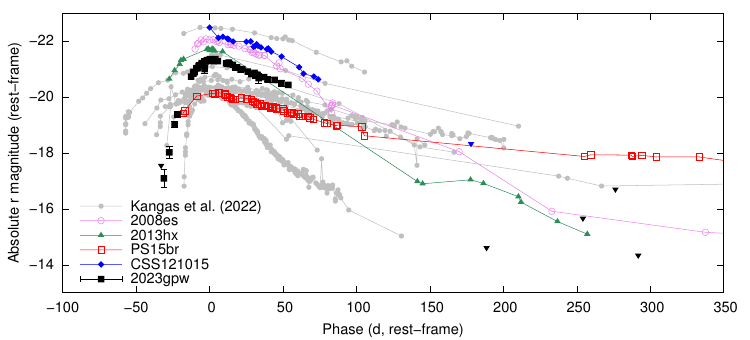}
\caption{Top panel: absolute $g$-band light curve of SN 2023gpw (black) compared to other SLSNe~II: we include the SLSNe~II presented by \citet{kangas22} (grey), the prototypical SN~2008es \citep[purple;][]{gezari09,miller09}, CSS121015 \citep{benetti14} and the two SLSNe~II presented by \citet{inserra18} (SN~2013hx in teal and PS15br in red). Bottom panel: same in the $r$ band. A pronounced steepening in the $r$-band light curve, such as that in SN~2023gpw at some point before $\sim+200$~d, is not observed in other SLSNe~II, though it is not excluded in some objects whose light curves do not continue this long. In both panels the zero epoch corresponds to the $g$-band peak.}
\label{fig:absLCcomp}
\end{figure*}

The observed photometric evolution of SN~2023gpw is relatively smooth, with no clear bumps or other conspicuous features. Simply based on the first detection at MJD\,=\,60051.3 and the $g$-band peak, the rise time is $\gtrsim$31~d. The ZTF photometry does not include a constraining non-detection before the first detection (the last two $r$-band non-detections at MJD\,=\,60049.3 and 60045.3 have upper limits of $\geq$20.4, brighter than the first detection with $r=20.78\pm0.32$~mag), but based on our GP fits, we can calculate the $g$-band rest-frame rise time from 10\% of the peak brightness, $t_\mathrm{rise,10\%} = 26.0^{+3.1}_{-2.9}$~d; from $1/e$ of the peak brightness, $t_\mathrm{rise,1/e} = 16.6^{+3.3}_{-3.0}$~d; and from half-maximum, $t_\mathrm{rise,0.5} = 15.4^{+3.2}_{-3.0}$~d. This rise is relatively fast compared to the SLSNe~II in \citet{kangas22}, especially compared to other SLSNe~II peaking at $<-21$~mag, and to the SLSNe~IIn in \citet{pessi24}. The decline rate after the peak depends on wavelength, with bluer bands declining faster. Optical decline rates measured using a linear fit between +20 and +54~d are $0.061\pm0.004$~mag~d$^{-1}$ in $u$, $0.034\pm0.001$~mag~d$^{-1}$ in $g$, $0.017\pm0.001$~mag~d$^{-1}$ in $r$ and $0.012\pm0.002$~mag~d$^{-1}$ in $i$.

After $\sim+55$~d, SN~2023gpw entered solar conjunction and no observations were performed in most filters until $+189$~d. In the meantime, only ATLAS photometry exists, extending to $+78$~d in the $c$ band and $+70$~d in the $o$ band. After this gap, no source is detected after host-galaxy subtraction. It is clear, however, that at the rate the $r$- and $i$-band light curves were declining when the gap began, they would not have reached the post-gap limits; in particular, in the $i$ band, there is a difference of almost 4~mag (see Fig. \ref{fig:absLC}). Some time during this gap, a drop of multiple magnitudes (depending on the filter) occurred. The drop had not yet begun at $+78$~d, the epoch of the last ATLAS detection. No further ATLAS data exist until after the gap, where only shallow limits (not deep enough to reach the host magnitude) are available. No clear broad lines from the SN are seen in a GTC/OSIRIS+ spectrum of the host galaxy taken at the position of the SN at +278~d, corroborating this result (see Sect. \ref{sec:specs}). Instead of an abrupt drop followed by a slower decline in SN~2023gpw, however, we cannot fully exclude a more gradually steepening light curve more similar to, for example, SN~2021irp, a SN~II with signs of interaction \citep{reynolds25}.

We show a comparison between the $g$- and $r$-band absolute light curves of SN~2023gpw and other SLSNe~II in Fig.~\ref{fig:absLCcomp}. We include the \citet{kangas22} sample, the prototypical SLSN~II event SN~2008es \citep{gezari09,miller09,kornpob19}, CSS121015:004244+132827 \citep[hereafter CSS121015;][]{benetti14} and the objects in \citet{inserra18}. For objects with $z>0.17$, the observed $r$ ($i$) band was converted to the rest-frame $g$ ($r$)~band. The $g$-band evolution of SN~2023gpw resembles that of SN~2008es and especially SN~2013hx, which exhibited a similar peak magnitude and decline, though SN~2023gpw rises to the peak faster than SN~2013hx and the rise of SN~2008es is not well constrained. Even at late times, these two SNe are within $\sim$0.5~mag of the limits we set for SN~2023gpw in the $g$ band (but a large gap in the light curve of SN~2013hx makes this less clear). By peak luminosity, SN~2023gpw is the sixth-brightest out of these 17 objects. The steepening light curve of SN~2023gpw during the solar conjunction is more pronounced in the $r$ band; such a drop is not observed among other SLSNe~II, but there are several objects where a drop around +100~d cannot be excluded due to a lack of data. Late-time observations of SN~2008es \citep{kornpob19} show that its light curve did not experience a steepening; instead, it declined relatively steadily until almost +300~d. 

\begin{figure}
\centering
\includegraphics[width=0.99\linewidth]{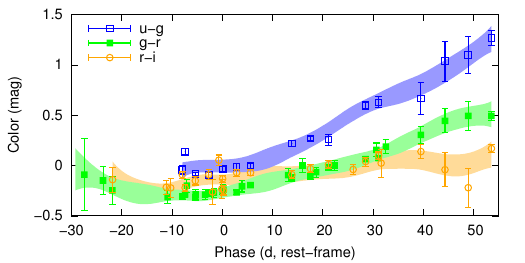}
\includegraphics[width=0.99\linewidth]{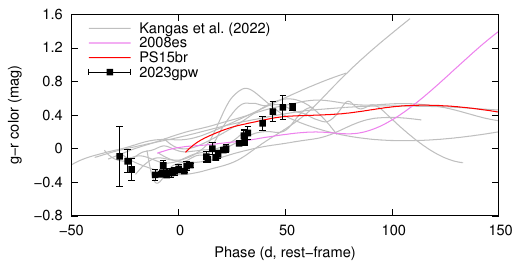}
\caption{Top panel: optical colors of SN~2023gpw, interpolated using a GP fit (shaded regions denote the $1\sigma$ uncertainty of this fit). The $g-r$ color seems to become bluer during the rise to the peak, but the uncertainties of the measured colors are also consistent with no evolution in $g-r$ until around the peak. Bottom panel: $g-r$ color of SN~2023gpw compared with other SLSNe~II \citep{miller09,inserra18,kangas22}, all corrected for Galactic extinction. The colors of comparison SNe have been GP-interpolated for visual clarity.}
\label{fig:colors}
\end{figure}

We show the color evolution of SN~2023gpw in Fig. \ref{fig:colors}, with the $g-r$ color also compared to other SLSNe~II (all corrected for Milky Way extinction). For the measured colors, we use data points within 0.1~d of each other in time; in the top panel, we also show the differences between the GP interpolations at each band as shaded regions, while in the bottom panel, the GP interpolations of comparison events are shown as lines. The uncertain early $g-r$ color is consistent with practically no change, as is the $r-i$ color. From around the peak, all measured colors ($u-g$, $g-r$ and $r-i$) become steadily redder over the rest of the light curve. The evolution of the $u-g$ color is the fastest, increasing from $\sim$0~mag at peak to $\sim$1.3~mag at 55~d. The $g-r$ color, between $\sim-0.3$~mag at $-10$~d and 0.5~mag at 50~d, is initially bluer than that of SN~2008es \citep{gezari09} or most SLSNe~II in the \citet{kangas22} sample, but also reddens faster than most SLSNe~II and eventually reaches a similar color.

\begin{figure}
\centering
\includegraphics[width=0.92\linewidth]{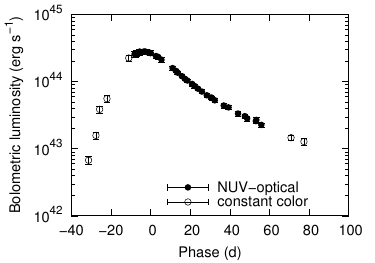}
\caption{\texttt{SuperBol} \citep{nicholl18} pseudo-bolometric light curve of SN~2023gpw. The values shown are based on either the full NUV-optical range (solid points), or only two or three bands and constant-color-based extrapolation of other bands (open points).}
\label{fig:bolLC}
\end{figure}

We used \texttt{SuperBol} \citep{nicholl18} to construct a pseudo-bolometric light curve. The $r$ band was used as the baseline, and other optical and NUV light curves were interpolated to the $r$-band epochs using a polynomial fit within \texttt{SuperBol}. Extrapolation of the early light curve before the first NUV and $i$ points was performed by assuming a constant color. Extrapolation to the end of the observations was done using $c$ as the baseline and assuming constant colors in the bands not observed between +54 and +78~d. NIR data were only available at one epoch and were not included. Blackbody-based bolometric corrections are performed in \texttt{SuperBol}, but we do not include them, as NUV data of SN~2023gpw deviate from a blackbody at most epochs (see Sect. \ref{sec:BB}); this also means that our pseudo-bolometric luminosities are lower limits for the total bolometric luminosity. The pseudo-bolometric light curve is displayed in Fig. \ref{fig:bolLC}. 

The pseudo-bolometric luminosity at its peak, at $\sim-5$~d, is $(2.8\pm0.3)\times 10^{44}$~erg~s$^{-1}$. It then declines by about a factor of 20 by the time of the solar conjunction. Before the conjunction the decline slows down slightly, but no bump or plateau is observed and the decline is smooth. Integrating over this light curve, we estimated the energy radiated by SN~2023gpw in the observed filters to be  ($8.91\pm0.21) \times 10^{50}$~erg, making this a lower limit for the total energy radiated across all wavelengths. \citet{kangas22} found similar or even greater energy requirements for the most luminous, 2008es-like SLSNe~II.

%--------------------------------------------------------------------
\subsection{Spectroscopy}
\label{sec:specs}

\begin{figure*}
\centering
\includegraphics[width=0.85\linewidth]{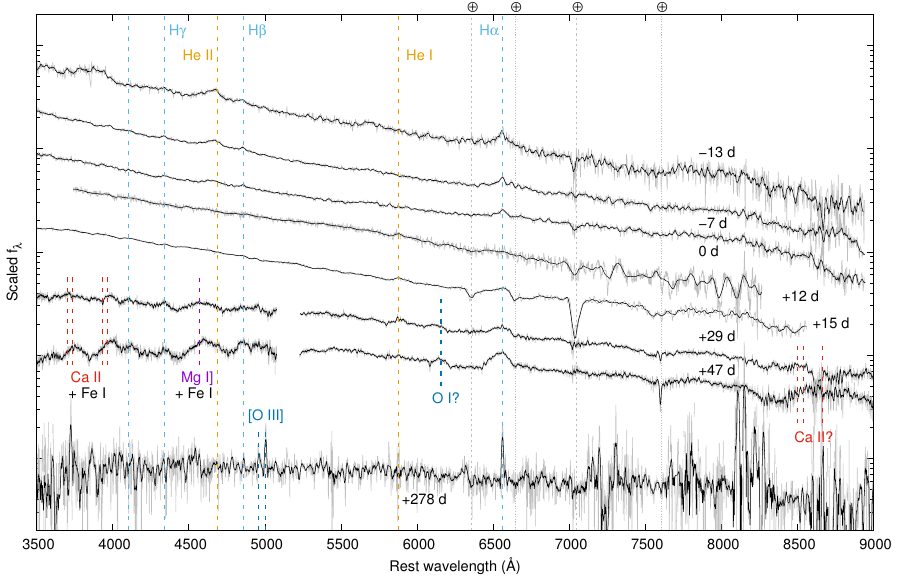}
\caption{Our moderate-resolution spectra of SN 2023gpw. Savitzky-Golay smoothing has been applied and the +12-d spectrum has been binned by a factor of 5; the original spectra are plotted in grey. The zero epoch corresponds to the $g$-band peak on MJD\,=\,60085.0. Telluric absorption is marked with the $\bigoplus$ symbol and vertical dotted lines. The +278-day spectrum has no clear SN lines; only narrow emission lines from the host galaxy can be seen. Extinction correction has not been applied. The public PESSTO spectrum at +15~d does not include telluric-line correction.}
\label{fig:spex}
\end{figure*}

\begin{figure}
\centering
\includegraphics[width=0.99\linewidth]{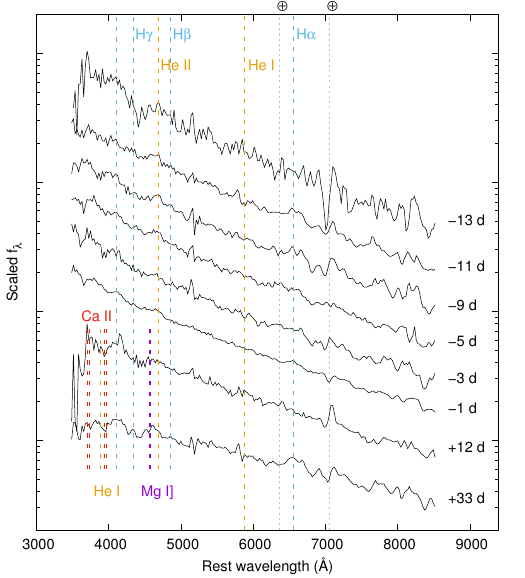}
\caption{Low-resolution SEDM spectra of SN 2023gpw. The zero epoch corresponds to the $g$-band peak on MJD\,=\,60085.0. Telluric absorption lines ($\bigoplus$) are marked in vertical dotted lines. The marked lines are based on the higher-resolution spectra; the Ca~{\sc ii} lines do not seem to match observed features here.}
\label{fig:sedms}
\end{figure}

\begin{figure}
\centering
\includegraphics[width=0.91\linewidth]{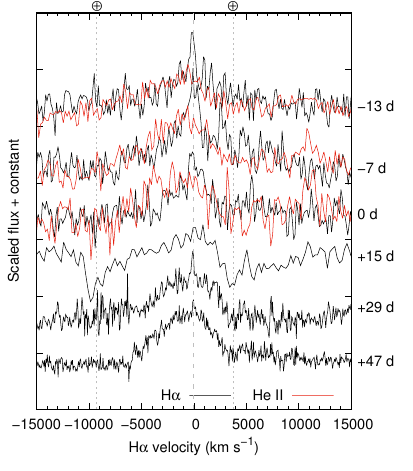}
\caption{Evolution of the H$\alpha$ (black) and He~{\sc ii}~$\lambda4686$ (red) line profiles in SN~2023gpw with the continuum subtracted on both sides. Telluric lines around H$\alpha$ ($\bigoplus$) are marked with dotted lines and the zero velocity with a dashed line. A weakening narrow component can be seen in the profile, while the wings of the broad component are disturbed by the telluric lines, but extend to several thousand km~s$^{-1}$ on both sides.}
\label{fig:hazoom}
\end{figure}

The spectra of SN~2023gpw used in this paper, except the low-resolution SEDM spectra, are shown in Fig. \ref{fig:spex}, including the classification spectrum \citep{kravtsov23}; the SEDM spectral sequence is shown in Fig. \ref{fig:sedms}. In the earliest spectra between $-13$~d and the peak, the visible features are Balmer lines and He~{\sc ii}~$\lambda4686$, of which H$\alpha$ and He~{\sc ii} are also visible in the SEDM spectra. Such `flash' features are common among normal SNe~II \citep{bruch21,bruch23}. The H$\beta$ and H$\gamma$ are faint in the ALFOSC spectrum at $-13$~d, but their presence is clear at $-7$~d and consistent with the $-13$~d spectrum. In Fig. \ref{fig:hazoom}, we show the H$\alpha$ profile evolution in particular; at early epochs this is compared to He~{\sc ii}~$\lambda4686$. The H$\alpha$ profile consists of a central narrow or intermediate-width component surrounded by broad wings, likely caused by electron scattering; the wings are quite clear at $-13$ and $-7$~d, but at peak, they have weakened. The full width at half-maximum (FWHM) of the narrow component is $\sim$1000~km~s$^{-1}$ in the ALFOSC spectra from $-13$~d to the peak. The velocity resolution at this wavelength is 680~km~s$^{-1}$, implying a marginally resolved line with an intrinsic width of $\sim$700~km~s$^{-1}$. The H$\beta$ line also exhibits a similar profile, but we see no clear narrow component in the He~{\sc ii}~$\lambda4686$ line, which can be fit using either a Lorentzian or a Gaussian function (residuals in both cases are dominated by noise), with a FWHM between 6400 and 7300~km~s$^{-1}$ depending on the spectrum and the function. By eye, the He~{\sc ii}~$\lambda4686$ line appears similar to the broad component of H$\alpha$ at $-13$~d. At all these epochs, the peak of He~{\sc ii}~$\lambda4686$ is blueshifted by between 1000 and 2000~km~s$^{-1}$, while the narrow component of H$\alpha$ is at zero velocity. 

All of these lines get weaker toward the peak, after which they disappear. Instead, we start to see a weak P~Cygni profile of He~{\sc i}~$\lambda5876$ and a broad H$\alpha$ line with a much weaker or non-existent narrow component. Neither kind of H$\alpha$ line is visible in the SEDM or INT/IDS spectra at +12~d, but the broad profile appears in later spectra starting at +15~d. The He~{\sc i}~$\lambda5876$ line, on the other hand, is already visible at +12~d. The velocity of the He~{\sc i}~$\lambda5876$ absorption minimum in the better-quality spectra at +15 and +29~d is between 4000 and 4500~km~s$^{-1}$. 

Telluric lines are strong in the PESSTO classification spectrum \citep{kravtsov23} and make it difficult to measure the H$\alpha$ velocity, but the profile in our DBSP spectra at +29 and +47~d can be fit using a Gaussian with a FWHM of $\sim$5500~km~s$^{-1}$. The profile at +29 and +47~d extends to roughly 5000~km~s$^{-1}$ on the blue side and slightly less on the red side, with the broad component blueshifted by $\sim$1000~km~s$^{-1}$, similarly to the early He~{\sc ii}. The H$\beta$ line is located close to the edge of the blue grating of DBSP and is difficult to fit. On top of the broad H$\alpha$, there is a weak unresolved (270~km~s$^{-1}$) component; this putative component is not much stronger than the noise level in the +47~d spectrum, but seems to be present at zero velocity at both epochs, similarly to the narrow component in the first two spectra. This weak narrow H$\alpha$ line is likely at least partially from the host galaxy. In any case, after the peak, the Balmer lines are dominated by the broad component.

The +47~d spectrum also shows apparent broad absorption lines in its red part. Atmospheric features likely contribute to this, as the +47-d spectrum was taken at a relatively high airmass during early twilight due to SN~2023gpw setting very early that night. The Ca~{\sc ii} NIR triplet is possibly present as well, but not clearly so. In the blue parts of the +29 and +47-d spectra, we see other broad lines appearing. These include the Mg~{\sc i}]~$\lambda4571$ and Ca~{\sc ii} doublets at $\lambda\lambda$3706,3736 and $\lambda\lambda$3934,3968. Unlike the Balmer lines, all of these seem to be redshifted by $\sim$2500~km~s$^{-1}$; it is also possible that these features have a contribution from nearby iron lines. The SEDM spectrum at +33~d does not seem to match the Ca~{\sc ii} identification, but it is also very noisy; the Mg~{\sc i}]~$\lambda4571$ line, on the other hand, is present. The He~{\sc i}~$\lambda5876$ line is no longer apparent at +47~d, but a broad feature possibly attributable to O~{\sc i}~$\lambda6157$ appears at +29 and +47~d. 

We used the late-time GTC spectrum on MJD\,=\,60386.2 (+278.4~d) to determine the redshift of SN~2023gpw. The spectrum is dominated by narrow host-galaxy emission, with no visible broad emission lines from the SN, except something that looks like an [O~{\sc i}]~$\lambda\lambda6300,6364$ feature; this emission, however, is only visible in one of the two OSIRIS exposures, and the 6364~\AA~component is missing, making it a likely artifact. No broad H$\alpha$, ubiquitous in late-time spectra of SLSNe~II \citep{kangas22}, is seen. We do, however, detect narrow lines of H$\beta$, [O~{\sc iii}]$~\lambda\lambda4959,5007$ and H$\alpha$, and using these lines we measured an average redshift of $z=0.0827\pm0.0002$. The strength of the [O~{\sc iii}]$~\lambda\lambda4959,5007$ doublet compared to H$\alpha$ suggests that the latter is also dominated by host galaxy emission and has little contribution from the SN. The lack of broad lines at such a late phase indicates that the observed emission is generally dominated by the host galaxy, which is consistent with the late-time non-detections in our photometry (Sect. \ref{sec:LC}). 

\begin{figure*}
\centering
\includegraphics[width=0.85\linewidth]{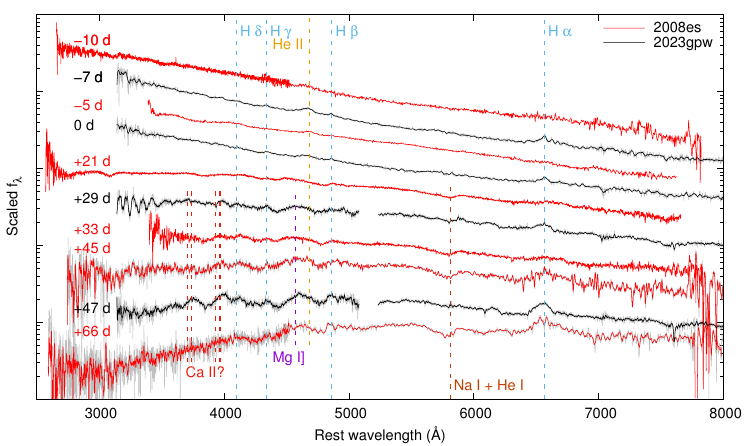}
\caption{Spectral evolution of SN~2023gpw (black) compared to the prototypical SN~2008es (red). Savitzky-Golay smoothing has been applied; the original spectra are plotted in grey. The zero epoch corresponds to the $g$-band peak.}
\label{fig:spexcomp1}
\end{figure*}

\begin{figure*}
\centering
\includegraphics[width=0.85\linewidth]{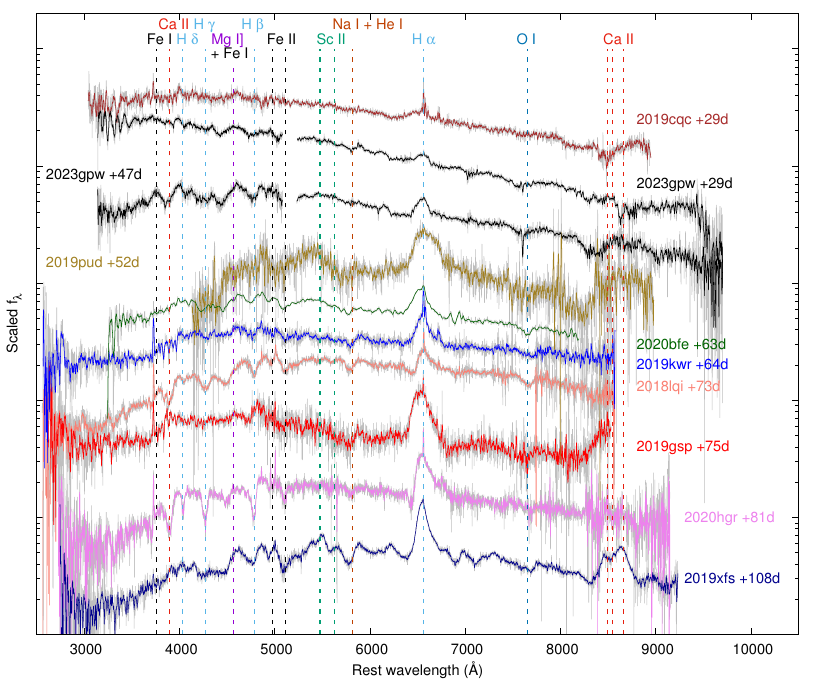}
\caption{Spectral evolution of SN~2023gpw (black) compared to SLSNe~II that resemble normal SNe~II more than they do SN~2008es \citep{kangas22}. Savitzky-Golay smoothing has been applied; the original spectra are plotted in grey. The zero epoch corresponds to the $g$-band peak.}
\label{fig:spexcomp2}
\end{figure*}

We show a comparison between SNe~2008es and 2023gpw in Fig. \ref{fig:spexcomp1}, and a comparison to those SLSNe~II that \citet{kangas22} considered more spectroscopically reminiscent of normal SNe~II in their sample in Fig. \ref{fig:spexcomp2}. Similarities can be seen to several objects; the strengths of the emission lines resemble those of SN~2008es, but the shape of the H$\alpha$ line is more similar to the latter group, albeit narrower than most. The spectra of the SLSNe~II resembling normal SNe~II almost all reach later epochs than what we have access to with SN~2023gpw, so we cannot say for sure whether its H$\alpha$ profile starts to resemble them later. The metal lines, if originating from Mg~{\sc i}]~$\lambda4571$ and Ca~{\sc ii}, would be similarly redshifted in the comparison objects, suggesting that there is, in fact, a large contribution from iron. In some cases, absorption components in P Cygni profiles are very clear, which we do not see in SN~2023gpw. Perhaps the best match to SN~2023gpw at +47~d among other SLSNe~II is SN~2020bfe at +63~d; however, no spectra of this SN exist between $-14$ and +63~d \citep{kangas22}. The same problem applies to other similar SLSNe~II. At $-14$~d, SN~2020bfe does show multi-component emission lines as well, but only those of hydrogen. Similarly to SN~2023gpw, SN~2020bfe also shows no H$\alpha$ absorption component and a weak He~{\sc i}~$\lambda5876$ (+Na~{\sc i}~$\lambda\lambda5890,5896$) feature. However, SN~2020bfe was over a magnitude fainter at peak and did not exhibit a steepening light curve during a follow-up extending to $\sim+220$~d.

\begin{figure}
\centering
\includegraphics[width=0.92\linewidth]{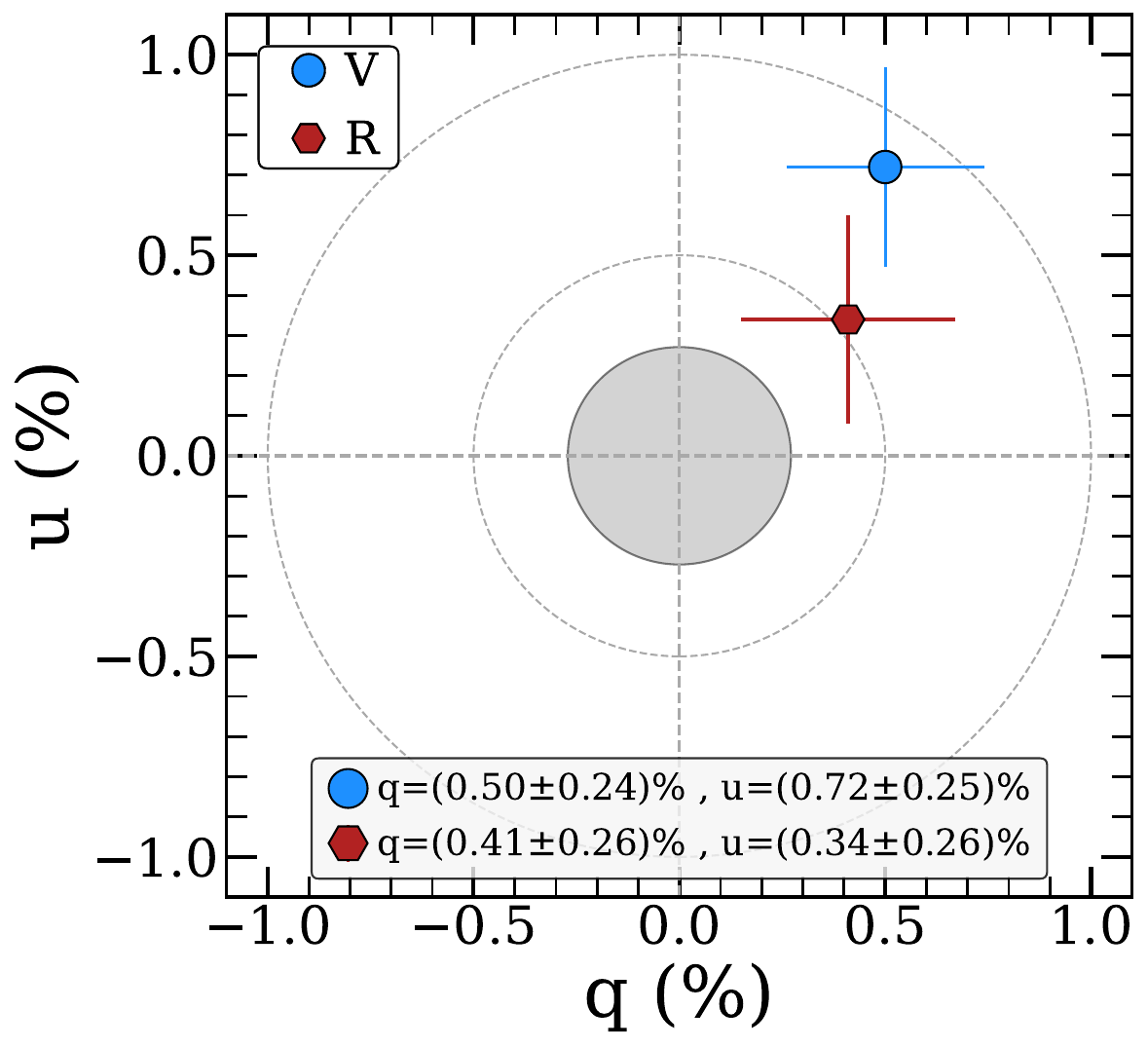}
\caption{Intensity-normalized Stokes $q$ and $u$ parameters of our polarimetry in the $V$ and $R$ bands on MJD\,=\,60081.0, corresponding to $-3.7$~d. The filled circle corresponds to our best ISP estimate (see text) and the dashed circles to 0.5\% and 1.0\% polarization. The two bands are consistent with each other within $1\sigma$ and higher than the ISP.}
\label{fig:stokesqu}
\end{figure}

\subsection{Polarization}
\label{sec:pol}

The intensity-normalized Stokes parameters ($q = Q/I$ and $u = U/I$,
where $Q$ and $U$ are the differences in flux with the electric field oscillating in two perpendicular directions, and $I$ is the total flux) were used to calculate the polarization degree (${p = \sqrt{q^{2} + u^{2}}}$) and the polarization angle (${\chi = 0.5\arctan(u/q)}$). All values of $p$ presented in this paper have been corrected for polarization bias \citep[e.g.][]{simmons85,wang97} following \citet{plaszczynski14}.

In the $V$ band, we measure ${q = (0.50\pm0.24)\%}$ and ${u = (0.72\pm0.25)\%}$, while in the $R$ band we measure ${q = (0.41\pm0.26)\%}$ and ${u = (0.34\pm0.26)\%}$. This leads to ${p = (0.84\pm0.25)\%}$ and ${\chi = (27.61\pm8.07)^\circ}$ in the $V$ band, and ${p = (0.47\pm0.26)\%}$ and ${\chi = (19.83\pm13.98)^\circ}$ in the $R$ band. The two bands are mutually consistent within $1\sigma$ in both polarization degree and angle, with an average degree of $p = (0.66\pm0.26)\%$. As the spectrum of the target is, at this epoch ($-3.7$~d), continuum-dominated, real differences between the bands from depolarization caused by lines are disfavored.

In order to properly isolate the intrinsic polarization, the effect of the interstellar polarization (ISP), introduced by dust grains in the line of sight, has to be estimated. Unfortunately, there are no stars with adequate S/N in the field of view of our ALFOSC observations. Another way to roughly estimate the Galactic ISP is as $9 \times E(B-V) \,\%$ \citep{serkowski75}, and for SN~2023gpw, this would be $\sim 0.27\,\%$. We also checked for polarization standard stars published in \citet{heiles00} that are close to the location of SN~2023gpw. We find one star within two degrees from SN~2023gpw ($\sim 1.24^\circ$) with a polarization value of ${p_\mathrm{ISP} = (0.26\pm0.07)\%}$, consistent with the above estimate. Based on the host galaxy (see Sect. \ref{sec:host}), there is no evidence for strong reddening, hence we neglect the ISP of the host. In Fig. \ref{fig:stokesqu} are shown the Stokes $q$ – $u$ planes for the imaging polarimetry, and the ISP estimate as a grey filled circle. The ISP, with a degree of $p_\mathrm{ISP} = 0.27\,\%$ and an unknown angle, cannot be directly removed from the observed polarization, but rather introduces an additional uncertainty of $p_\mathrm{ISP}/\sqrt{2} \approx 0.20$\,\%, for a total of $p = (0.66\pm0.33)\%$, only 2$\sigma$ above zero polarization.

Another thing to consider when dealing with nuclear transients is the effect of dilution by unpolarized light from the host galaxy. However, in the case of SN~2023gpw, the host flux is much fainter than the transient flux during our epoch of observation (see Sect. \ref{sec:host}), hence the effect is negligible. Considering all of the above, we conclude that SN~2023gpw is likely intrinsically polarized, though its low significance makes it impossible to be certain. This polarization would indicate asphericity in SN~2023gpw during the epochs before and/or around the peak. 

%--------------------------------------------------------------------
\section{Modeling}
\label{sec:models}

\subsection{Blackbody fits}
\label{sec:BB}

\begin{figure}
\centering
\includegraphics[width=0.99\linewidth]{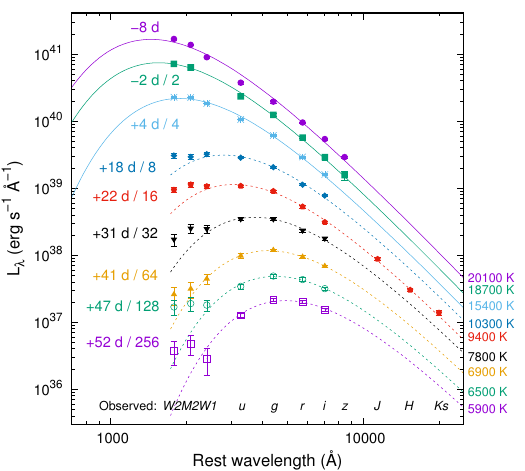}
\caption{Blackbody fits (lines) to the observed SEDs of SN~2023gpw (points). An excess over the blackbody function appears and gradually strengthens after the peak. The $W2$ and $M2$ bands have been ignored in the fits denoted with dashed lines; the $W1$ band is roughly compatible with the optical data throughout the evolution.}
\label{fig:bbfits}
\end{figure}

We performed blackbody fits to the spectral energy distributions (SEDs) of SN~2023gpw using our NUV, optical and NIR photometry. This was done at each NUV epoch; in the optical, we used our GP-interpolated light curves (see Sect. \ref{sec:LC}) to match these epochs. The NIR data points cannot be interpolated as we only have one epoch, but on MJD\,=\,60109.9 they are close to the NUV epoch on MJD\,=\,60108.6 and were used in the fit at this epoch. Starting at $+18$~d, an excess over the blackbody function is immediately apparent in the NUV, and the $W2$ and $M2$ bands were ignored in the fits at $+18$~d and afterwards. These fits are shown in Fig. \ref{fig:bbfits}. Earlier, the NUV points match well with the blackbody (though the $u$-band point at $-8$~d may deviate from the other points). The flattening NUV spectrum does not strongly affect the $W1$ band, which is compatible with the same blackbody as the optical and NIR data. Before the start of our NUV follow-up, we only have the ZTF $gri$-band data available, and the resulting blackbody fits are less robust. 

\begin{figure}
\centering
\includegraphics[width=0.99\linewidth]{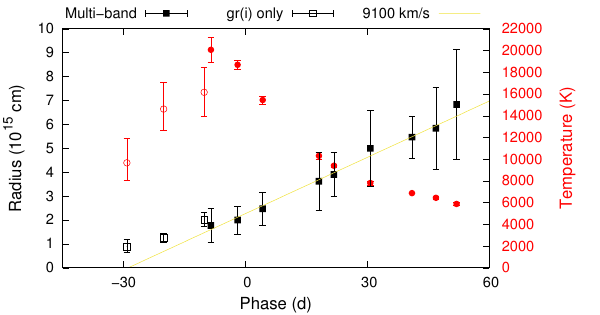}
\caption{Evolution of the blackbody temperature (red) and radius (black) of SN~2023gpw. The empty symbols are based on less reliable fits using only the $gr(i)$-band data. The yellow line corresponds to the best-fit expansion velocity using the solid points only; this line reaches zero radius at $-$29~d, close to the first observed epoch at $-$31~d.}
\label{fig:bbparams}
\end{figure}

We show our blackbody parameters, the temperature and the radius, in Fig. \ref{fig:bbparams}, including the fits based only on the $gr(i)$ bands. The blackbody radius increases linearly over time; a fit to this expansion (ignoring the epochs before NUV observations) results in a photospheric velocity of $v_{ph} = 9080\pm430$~km~s$^{-1}$, significantly higher than the spectroscopic velocities measured from the hydrogen and helium lines. Extrapolating the best-fit velocity to earlier epochs, we obtain an explosion date of $-28.9\pm1.6$~d, which is close to the discovery epoch at $-31.1$~d and hints that the SN was indeed discovered quite close to the explosion (or, in principle, that the expansion speed increased early on). The temperature at early times is not particularly reliable as we only have two or three bands available, but it seems to increase during the rise based on the $g-r$ color. At peak, the temperature is $\sim$20\,000~K \citep[similar to the hottest SNe~II at peak, but somewhat later due to the slower rise; see][]{faran18}. This then declines to $\sim$10\,000~K by $\sim$+20~d and, afterward, declines more slowly to around 6000~K at +50~d. 

\subsection{Light-curve fits}
\label{sec:mosfit}

We used the publicly available code Modular Open Source Fitter for Transients \citep[\texttt{MOSFiT}\footnote{\url{https://mosfit.readthedocs.io/en/latest/index.html}};][]{guillochon18} to fit the multi-band light curve. We included the radioactive $^{56}$Ni decay model \citep{arnett82,nadyozhin94}, labeled \texttt{default}; a CSI model \citep{chatz13,villar17,jiang20}, labeled \texttt{csm}; a fallback accretion model \citep{moriya18b}, labeled \texttt{fallback}; and a magnetar central engine model \citep{kb10}, labeled \texttt{magnetar}, included in the code. 
The $W2$ and $M2$ light curve data were ignored in our fits, as a UV excess is not reproduced in these models. The \texttt{slsn} model \citep{nicholl17} is a modified version of the \texttt{magnetar} model with constraints for more plausible physical parameters, limiting the neutrino energy to $\sim10^{51}$~erg and cutting out models that become nebular before 100~d. However, it also includes a modified blackbody with absorption below 3000~\AA. This is commonly observed in SLSNe~I but not in SN~2023gpw, where we would also need to ignore the $W1$ band. Hence, we instead modified the \texttt{magnetar} model file to include the same constraints.

These models were fitted using the dynamic nested sampling package \texttt{dynesty} \citep{speagle20}.\footnote{\url{https://dynesty.readthedocs.io/en/latest/}} We ran each fitting process until convergence, which typically took between 15\,000 and 25\,000 iterations. The quality of the fit is quantified by a likelihood score returned by \texttt{MOSFiT}, corresponding to the logarithm of the Bayesian evidence $Z$.

\begin{table}
\centering
\caption{Free and fixed parameters used in each \texttt{MOSFiT} model fit to the second peak, with prior distributions indicated. }
\label{tab:priors}
\begin{tabular}{ccc}
\hline
Parameter & Distribution & Range \\
\hline
\multicolumn{3}{c}{Common parameters} \\
\hline
$n_{H,\mathrm{host}}$ & [$10^{16}$ : $2\times10^{21}$] cm$^{-2}$ & log-uniform \\
$t_{\mathrm{expl}}$ & [$-$50 : 0] d & uniform \\
$M_\mathrm{ej}$ & [0.1 : 100]~$\mathrm{M}_\odot$ & log-uniform \\
$v_\mathrm{ej}$ & [5000 : 15\,000]~km~s$^{-1}$ & uniform \\
\hline
\multicolumn{3}{c}{$^{56}$Ni model} \\
\hline
$T_\mathrm{min}$ & [1000 : 50\,000] K & log-uniform \\
$f_{\mathrm{Ni}}$ & [$10^{-3}$ : 1.0] & log-uniform \\
$\kappa$ & 0.34~cm$^2$~g$^{-1}$ & fixed \\
$\kappa_\gamma$ & [0.1 : $10^4$]~cm$^2$~g$^{-1}$ & log-uniform \\
\hline
\multicolumn{3}{c}{Magnetar model} \\
\hline
$\kappa$ & 0.34~cm$^2$~g$^{-1}$ & fixed \\
$\kappa_\gamma$ & [0.1 : $10^4$]~cm$^2$~g$^{-1}$ & log-uniform \\
$P_{\mathrm{spin}}$ & [1 : 20] ms & uniform \\
$B_\perp$ & [0.1 : 50]~$\times10^{14}$ G & log-uniform \\
$M_\mathrm{NS}$ & [1.0 : 2.5]~$\mathrm{M}_\odot$ & uniform \\
$\theta_\mathrm{PB}$ & [0 : $\pi/2$] rad & uniform \\
\hline
\multicolumn{3}{c}{CSI model} \\
\hline
$T_\mathrm{min}$ & [1000 : 50\,000] K & log-uniform \\
$n$ & 12 & fixed \\
$\delta$ & 1 & fixed \\
$s$ & 0 or 2 & fixed \\
$R_0$ & [0.1 : 250] AU & log-uniform \\
$M_\mathrm{CSM}$ & [0.1 : 100]~$\mathrm{M}_\odot$ & log-uniform \\
$\rho$ & [$10^{-15}$ : $10^{-6}$]~g~cm$^{-3}$ & log-uniform \\
\hline
\multicolumn{3}{c}{Fallback model} \\
\hline
$T_\mathrm{min}$ & [1000 : 50\,000] K & log-uniform \\
$\kappa$ & 0.34~cm$^2$~g$^{-1}$ & fixed \\
$t_{tr}$ & [$10^{-4}$:100] d & log-uniform \\
$L_1$ & [$10^{50}$:$10^{57}$] erg s$^{-1}$ & log-uniform \\
\hline
\end{tabular}
\end{table}
 
We set simple, broad, uniform or log-uniform priors for each free parameter, summarized in Table \ref{tab:priors} for each model. Based on the lack of narrow Na~{\sc i}~D absorption lines in the spectra, we set an upper limit for the host galaxy extinction, $A_{V,\mathrm{host}} \leq 1$~mag. Host extinction itself is not a parameter in \texttt{MOSFiT}, but a related quantity, the column density of neutral hydrogen, $n_{H,\mathrm{host}}$, is. We therefore set the upper limit as $n_{H,\mathrm{host}} \leq 2\times10^{21}$~cm$^{-2}$ based on \citet{guver09}, similarly to \citet{kangas22}. 

Other parameters common to all models include the time from explosion to observations $t_\mathrm{expl}$, the minimum temperature $T_\mathrm{min}$ and the ejecta mass $M_\mathrm{ej}$. We fixed the (Thomson scattering) opacity $\kappa$ at 0.34~cm$^2$~g$^{-1}$, a typical value for hydrogen-rich ejecta and close to the result of \citet{nagy18}. This parameter is included in all models used here except the CSI model. 

The $^{56}$Ni model also includes the nickel fraction in the ejecta $f_{\mathrm{Ni}}$ and the opacity to $\gamma$-rays $\kappa_\gamma$. The magnetar model includes the spin period $P_\mathrm{spin}$, the magnetic field perpendicular to the spin axis $B_\perp$, the neutron star mass $M_\mathrm{NS}$ and the angle between the magnetic field and spin axis $\theta_\mathrm{PB}$. In the CSI model, we assumed a hydrogen-rich progenitor, but not necessarily an extended envelope. Thus the minimum inner radius of the CSM, $R_0$, was set at 0.1 AU ($\sim20 \,R_\odot$), roughly half the radius of the blue supergiant progenitor of SN 1987A \citep{podsiadlowski92}, but larger than a Wolf-Rayet progenitor of a stripped-envelope SN. From the disappearance of the `flash' emission lines before +12~d, $\gtrsim$43~d after the explosion, and the ejecta velocity of 9100~km~s$^{-1}$, we can estimate that some part of the CSM has been encountered by the ejecta before $3.4 \times 10^{15}$~cm, corresponding to $\lesssim230$~AU; conservatively, we set the upper limit for $R_0$ at 250~AU. The density of the CSM at $R_0$, $\rho$, is also a free parameter. We fixed the density power-law parameters in the inner ($\rho_{ej} \propto r^{-\delta}$) and outer ($\rho_{ej} \propto r^{-n}$) ejecta, $\delta=1$ and $n=12$, respectively -- normal values in hydrogen-rich ejecta \citep[][]{chatz13}. The density power law of the CSM ($\rho_{\mathrm{CSM}} \propto r^{-s}$) was fixed at either $s=0$ or $s=2$, respectively corresponding to a constant-density or wind-like CSM. Finally, the fallback model includes the linear-to-power-law accretion transition time $t_{tr}$ and the accretion luminosity at the transition time $L_1$. 

\begin{figure*}
\centering
\includegraphics[width=0.47\linewidth]{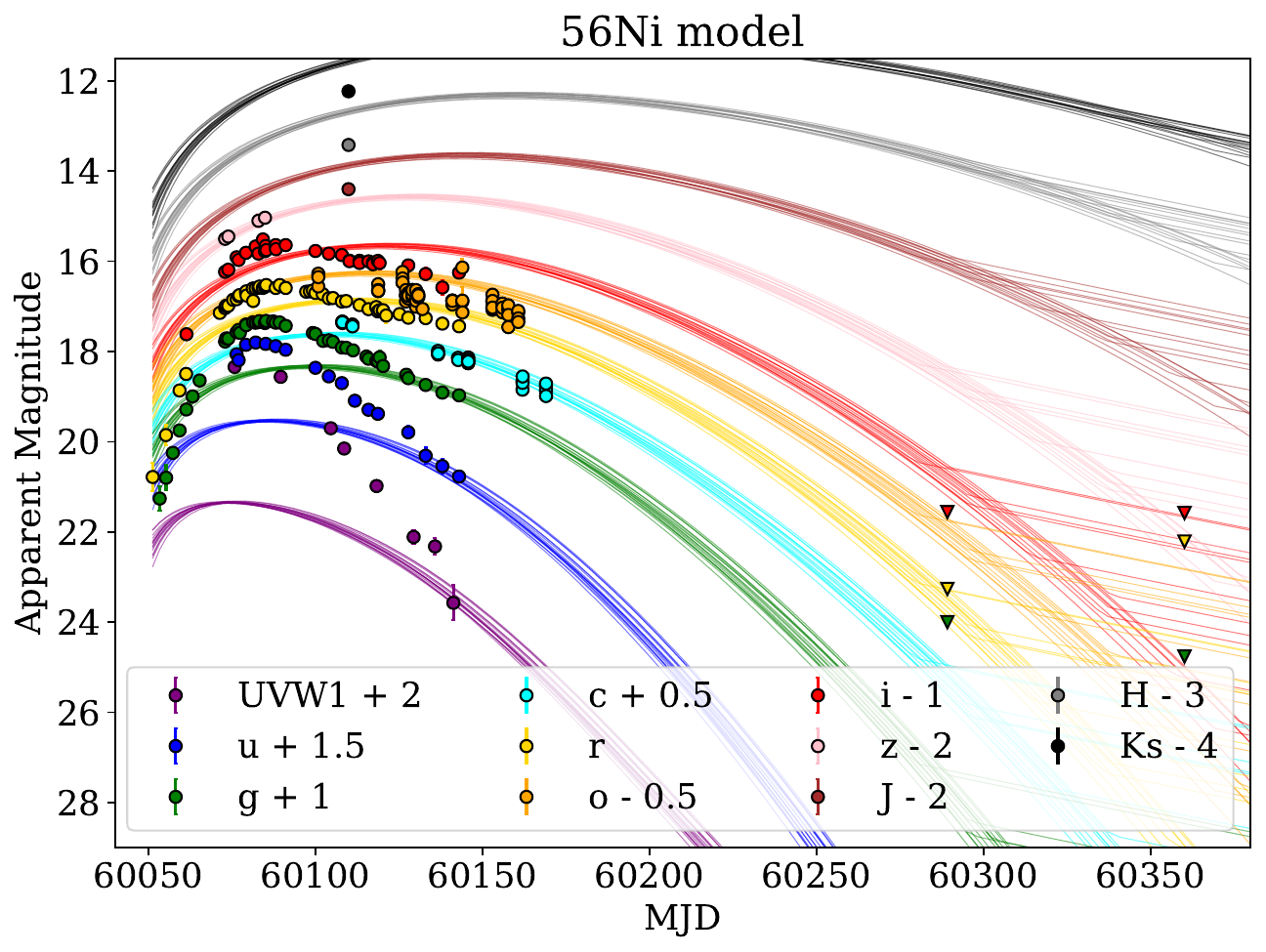}
\includegraphics[width=0.47\linewidth]{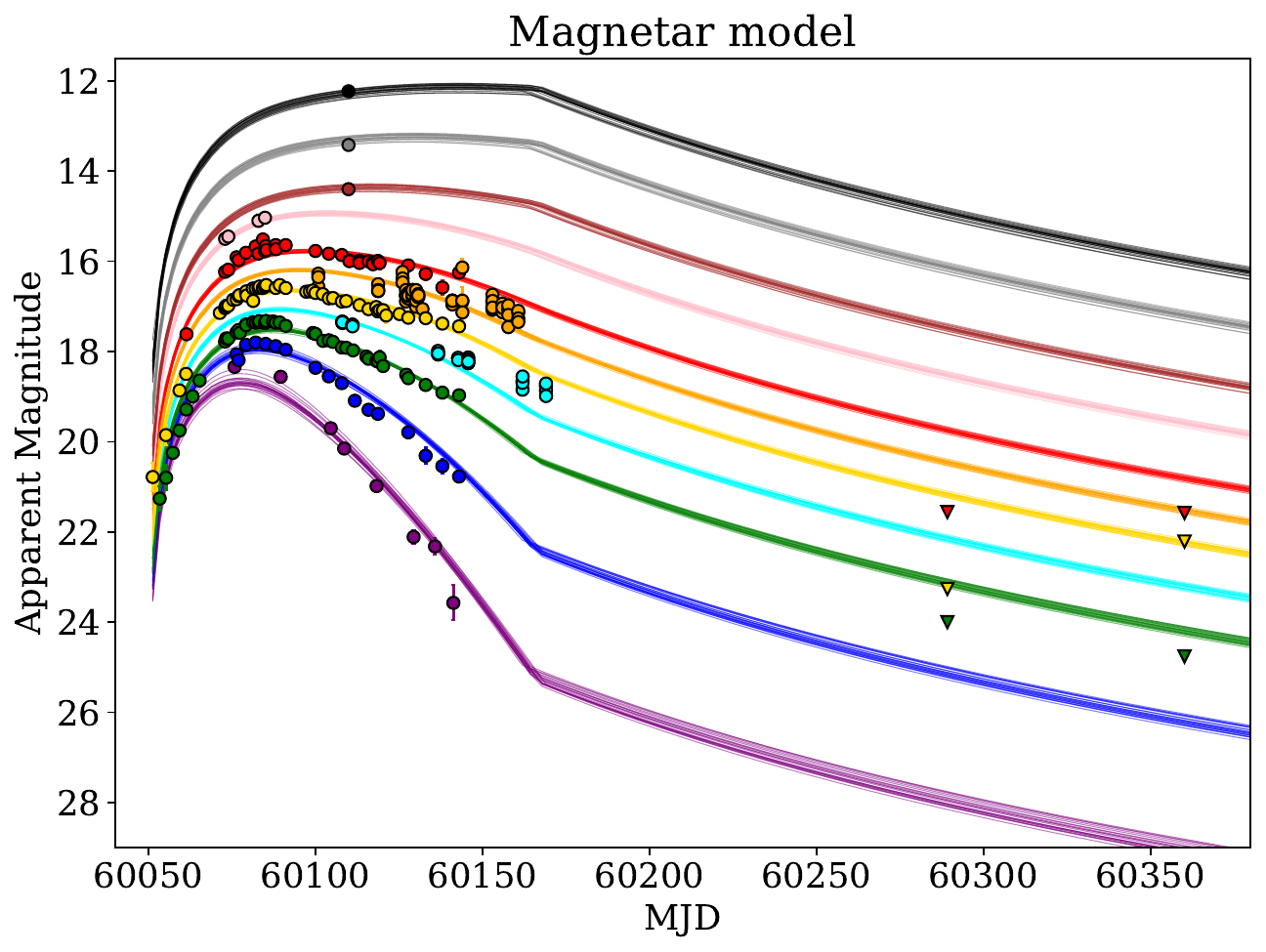}
\includegraphics[width=0.47\linewidth]{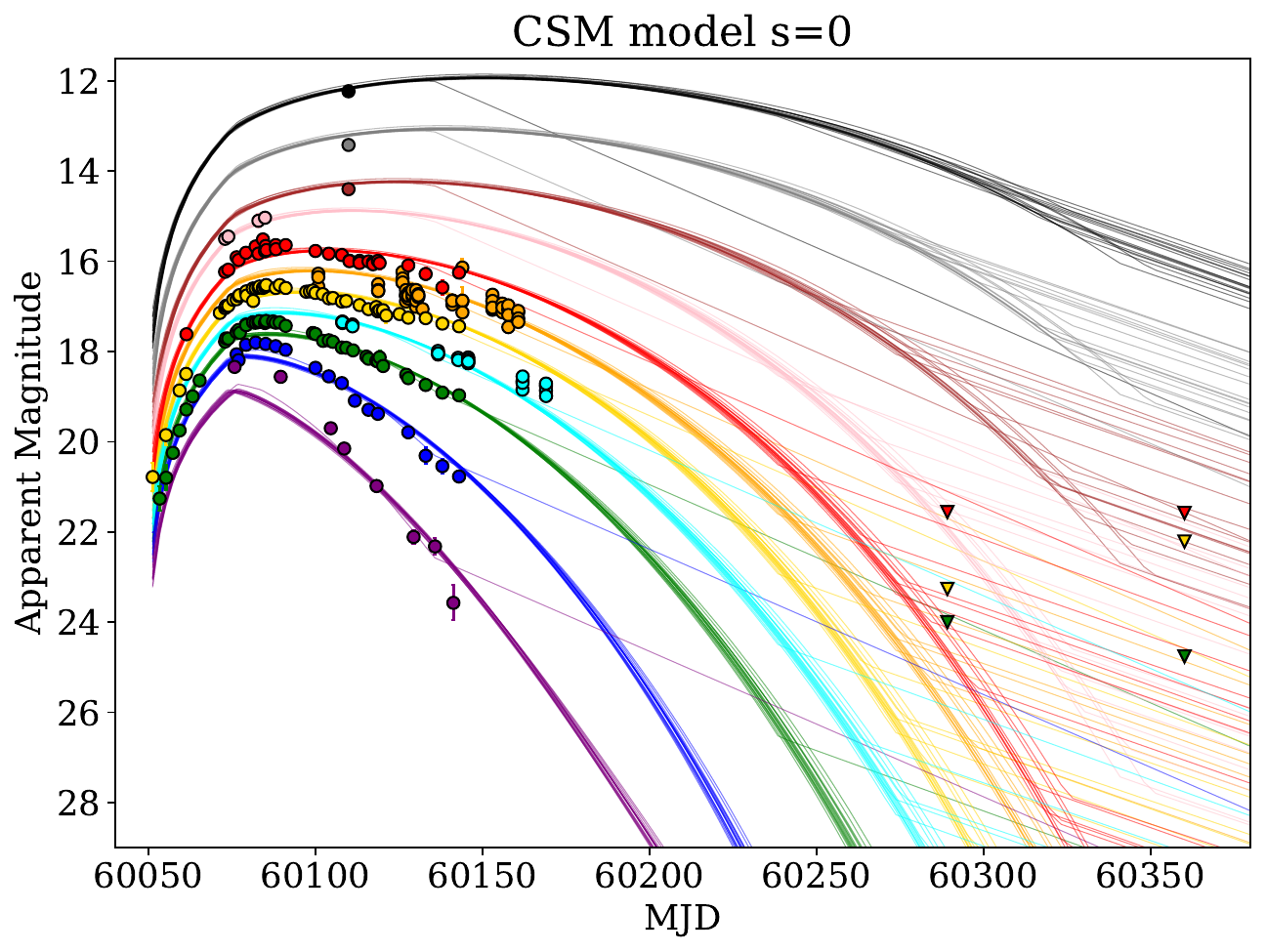}
\includegraphics[width=0.47\linewidth]{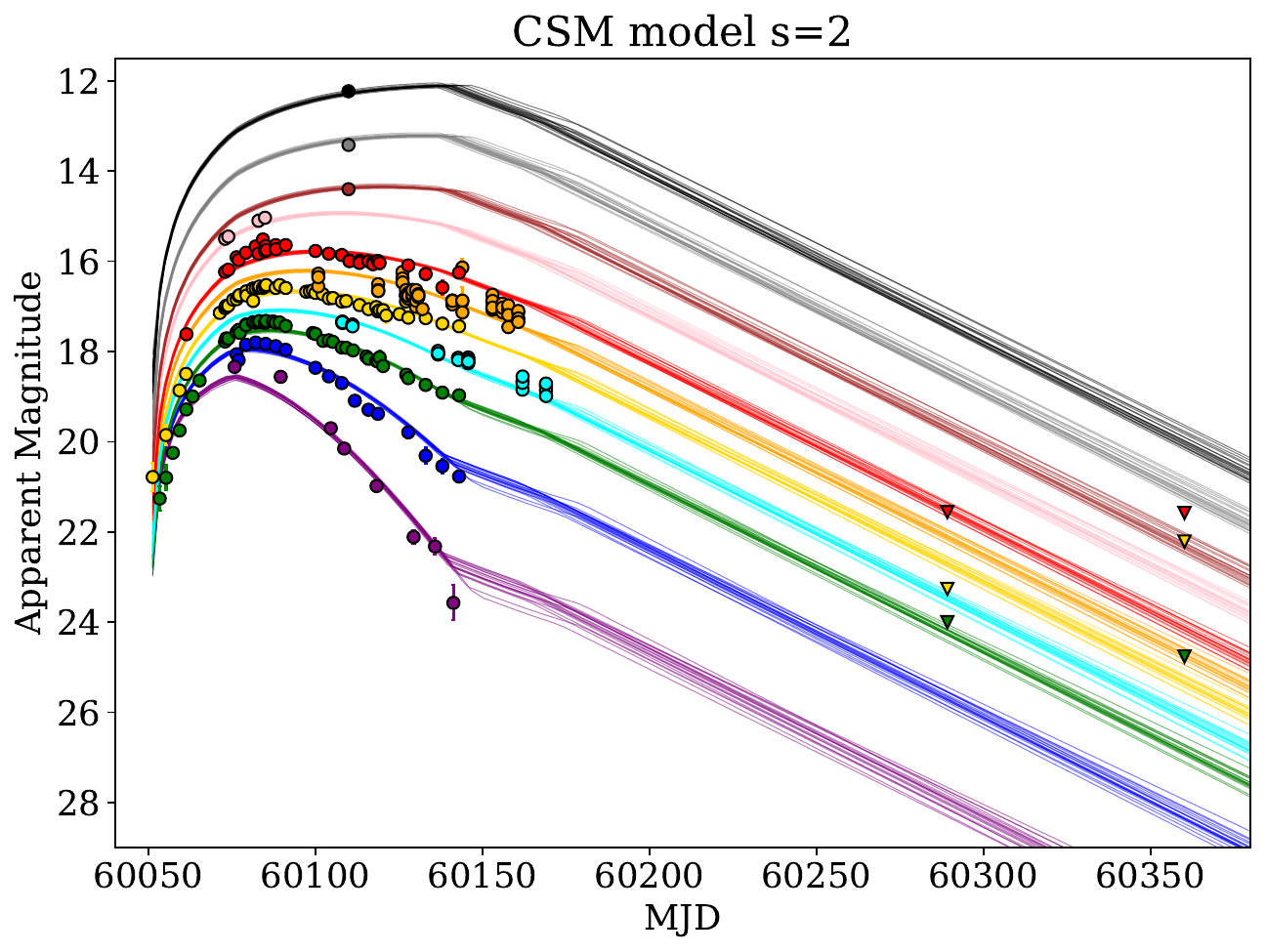}
\includegraphics[width=0.47\linewidth]{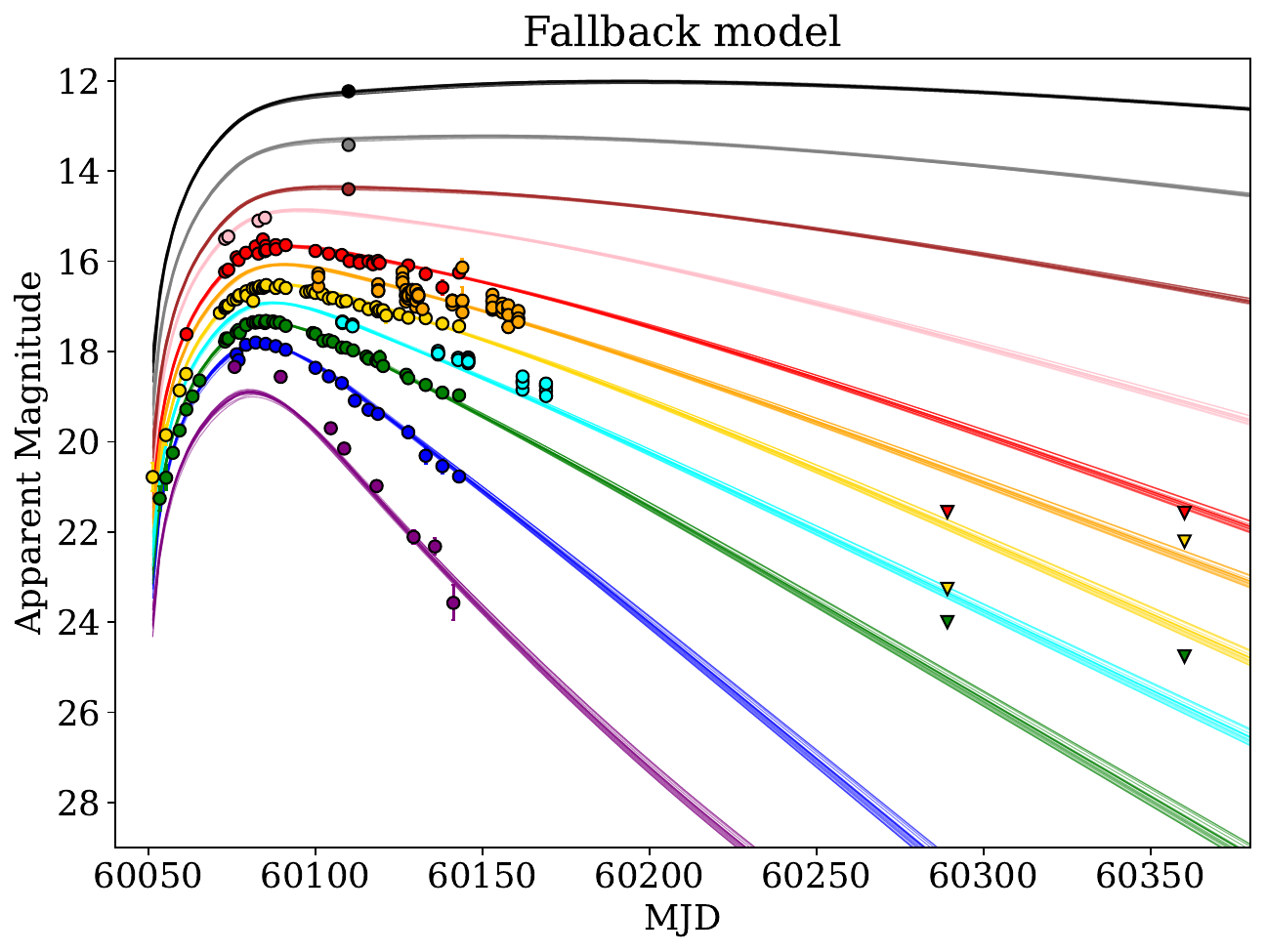}
\caption{\texttt{MOSFiT} light curve fits to SN~2023gpw using the different models described in the text. Late-time upper limits are represented as points where the $3\sigma$ uncertainty corresponds to the measured limit. All models, but especially \texttt{magnetar} and \texttt{fallback}, have trouble reproducing the drop in brightness during the gap. The best fit by eye and score is the CSM model with $s=0$. }
\label{fig:mosfits}
\end{figure*}

In total, the \texttt{default} and \texttt{fallback} models have 8 free parameters, the \texttt{csm} model has 9, and the \texttt{magnetar} model has 10. This includes a nuisance parameter $\sigma$, which describes the added variance required to match the model being fitted. The outputs of the \texttt{MOSFiT} modeling are shown in Fig. \ref{fig:mosfits}. The median posterior values of each free physical parameter, with the 16th and 84th percentile values as uncertainties, along with the Bayesian evidence score of each model, are listed in Table \ref{tab:mosfitres}. The corner plots of the fits, which also include $\sigma$, can be found in the appendices in Figs. \ref{fig:cornerni56} to \ref{fig:cornerFB}.

\begin{table*}
\centering
\caption{Median posterior values and uncertainties of free parameters in each \texttt{MOSFiT} model fit, and the score of each model.} 
\label{tab:mosfitres}
\begin{tabular}{lccccc}
\hline
Parameter & $^{56}$Ni decay & Magnetar & CSI ($s=0$) & CSI ($s=2$) & Fallback \\
\hline
log $n_{H,\mathrm{host}}$ (cm$^{-2}$) & $17.4^{+1.1}_{-1.0}$ & $20.63^{+0.11}_{-0.19}$ & $17.6^{+1.0}_{-1.1}$ & $18.9^{+1.2}_{-1.8}$ & $17.6\pm1.1$ \\
$t_{\mathrm{expl}}$ (d) & $-10\pm2$ & $-1.8^{+0.2}_{-0.3}$ & $-3.1^{+0.4}_{-0.5}$ & $-1.1\pm0.2$ & $-2.1\pm0.3$\\
log $T_\mathrm{min}$ (K) & $3.16\pm0.11$ & - & $3.2\pm0.2$ & $3.80\pm0.01$ & $3.1\pm0.1$\\
log $M_\mathrm{ej}$ ($\mathrm{M}_\odot$) & $1.00\pm0.02$ & $-0.08^{+0.06}_{-0.07}$ & $0.83^{+0.06}_{-0.07}$ & $1.48^{+0.04}_{-0.02}$ & $0.08^{+0.05}_{-0.06}$ \\
$v_\mathrm{ej}$ (km~s$^{-1}$) & $14800^{+200}_{-400}$ & $8200\pm300$ & $9500\pm200$ & $8900\pm200$ & $8700^{+300}_{-200}$ \\
log $f_{\mathrm{Ni}}$ & $-0.002^{+0.002}_{-0.003}$ & - & - & - & - \\
log $\kappa_\gamma$ (cm$^2$~g$^{-1}$) & $-0.97^{+0.06}_{-0.03}$ & $-0.94^{+0.06}_{-0.05}$ & - & - & - \\
$P_{\mathrm{spin}}$ (ms) & - & $2.7^{+0.7}_{-0.6}$ & - & - & -\\
log $B_\perp$ ($10^{14}$ G) & - & $0.11\pm0.17$ & - & - & -\\
$M_\mathrm{NS}$ ($\mathrm{M}_\odot$) & - & $1.7\pm0.5$ & - & - & -\\
$\theta_\mathrm{PB}$ (rad) & - & $0.6\pm0.3$ & - & - & -\\
log $R_0$ (AU) & - & - & $0.5\pm1.1$ & $0.9\pm0.5$ & -\\
log $M_\mathrm{CSM}$ ($\mathrm{M}_\odot$) & - & - & $0.60\pm0.02$ & $0.50^{+0.03}_{-0.02}$ & -\\
log $\rho$ (g cm$^{-3}$) & - & - & $-12.59^{+0.04}_{-0.03}$ & $-10.7\pm0.9$ & -\\
log $t_{tr}$ & - & - & - & - & $1.39\pm0.03$ \\
log $L_1$ & - & - & - & - & $55.05\pm0.01$ \\
\hline
Score (log $Z$) & -41.5 & 147.1 & 250.7 & 249.4 & 237.8 \\
\hline 
\end{tabular}
\end{table*}

It is clear that with these constraints, the \texttt{default} model cannot account for the rise and decline times of the light curve simultaneously with its luminosity. This is unsurprising, as the $^{56}$Ni decay model is incompatible with SLSNe II in general \citep{kangas22}. The \texttt{magnetar} and \texttt{fallback} models can somewhat replicate the early light curve, but both models result in a power-law decline at late times ($t^{-2}$ or $t^{-5/3}$, respectively), which over-predicts the flux after the solar conjunction. In the \texttt{csm} model with $s=2$ the late-time behavior is still off, but somewhat less so. Meanwhile, the $s=0$ case looks quite different and does produce a sufficient drop in brightness during the gap -- it is also the best fit by \texttt{MOSFiT} score, albeit very narrowly. In the $s=2$ model, the (hydrogen-rich) SN ejecta, with a large mass of $30^{+3}_{-2}$~M$_\odot$, would interact with a CSM of $3.2^{+0.3}_{-0.2}$~M$_\odot$, while in the $s=0$ case, we obtain a smaller ejecta mass of $6.8\pm1.0$~M$_\odot$ with a larger CSM mass, $4.0\pm0.2$~M$_\odot$. All these models except \texttt{default} return explosion times between $\sim$1 and $\sim$3 days before the first detection of SN~2023gpw, indicating that it was caught soon after the explosion.

%--------------------------------------------------------------------
\subsection{Host galaxy}
\label{sec:host}

\begin{figure}
\centering
\includegraphics[width=0.99\linewidth]{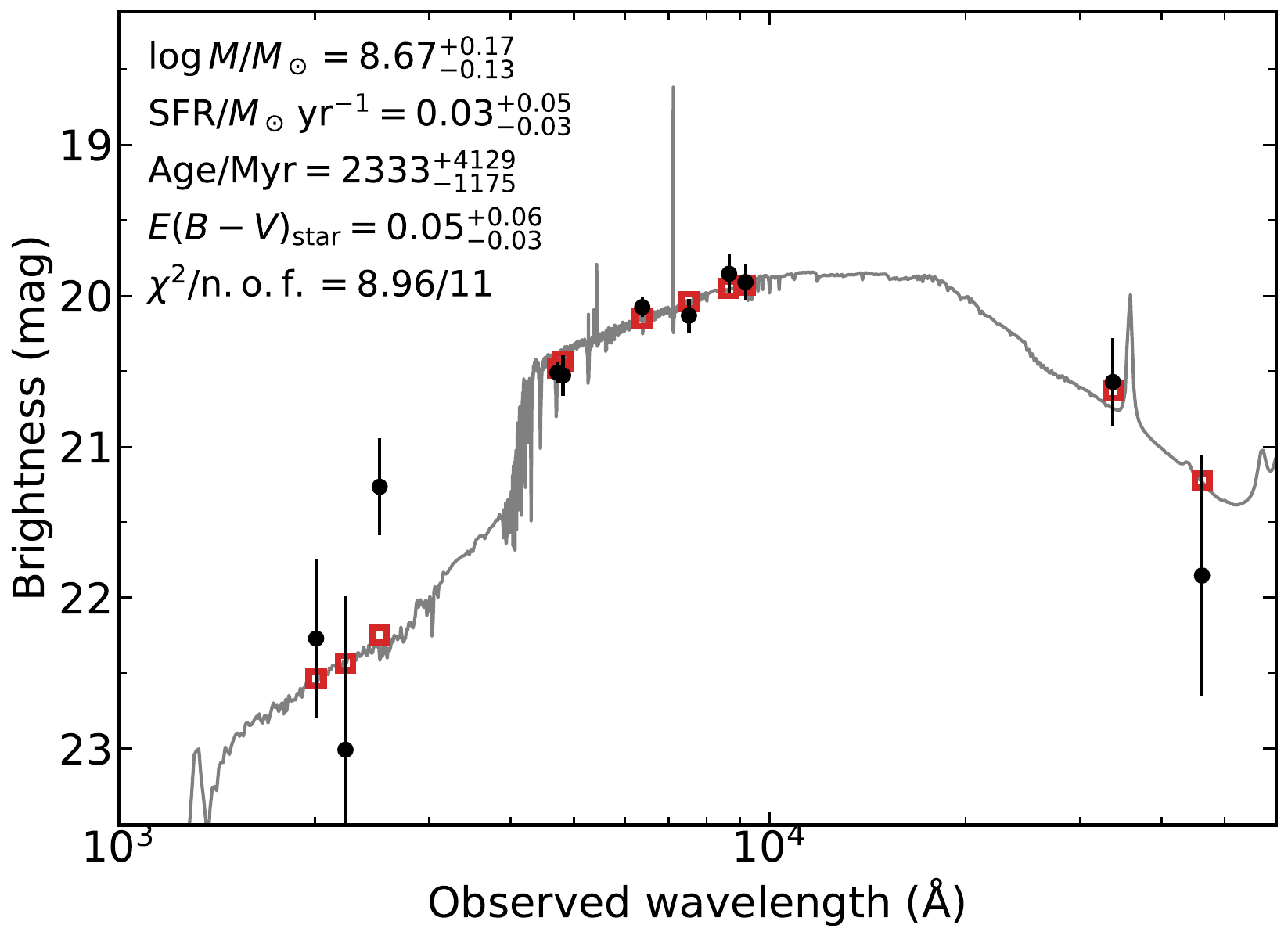}
\caption{SED of the host galaxy of SN 2023gpw from 1000 to 60\,000 Å (black points), and the best-fitting \texttt{Prospector} model (solid line). The red squares represent the model-predicted magnitudes; some deviation can be seen in the NUV. The fitting parameters are shown in the upper-left corner. The abbreviation "n.o.f." stands for the number of filters.
}
\label{fig:hostfit}
\end{figure}

To quantify the proximity of the transient to the nucleus of the host galaxy, we first aligned our late-time $r$-band images dominated by the host galaxy at sub-pixel precision using the \texttt{Image Reduction and Analysis Facility} (\texttt{IRAF}\footnote{\url{https://iraf-community.github.io/}}) tasks \texttt{geomap} and \texttt{geotran}, then added these together to create a stacked host galaxy image. We then performed a Markov Chain Monte Carlo fit of a two-dimensional S\'{e}rsic profile to this image using the \texttt{Sersic2D} function in \texttt{astropy}\footnote{\url{https://www.astropy.org/index.html}} and the \texttt{emcee}\footnote{\url{https://emcee.readthedocs.io/en/stable/}} package \citep{foreman13} to find the centroid in pixel space (with the centroid, S\'{e}rsic index, ellipticity and position angle as free parameters). All our $r$-band images of the SN from May 2023 (7 images), around the peak of the light curve, were aligned to the host galaxy image and the centroid of the SN measured in each using a Moffat profile fit in \texttt{IRAF}. The average difference in the centroids is $0\farcs81\pm0\farcs03$, which corresponds to $1.5\pm0.1$~kpc at the distance of $377.3\pm1.0$~Mpc.

To characterize the host galaxy itself, we retrieved public science-ready coadded pre-explosion images from the DESI Legacy Imaging Surveys \citep[Legacy Surveys, LS;][]{Dey2018a} Data Release (DR) 10, the PS1 DR 1 \citep{Chambers2016a}, and preprocessed \textit{WISE} images \citep{Wright2010a} from the unWISE archive \citep{Lang2014a}.\footnote{\href{http://unwise.me}{http://unwise.me}} The unWISE images are based on the public \textit{WISE} data and include images from the ongoing NEOWISE-Reactivation mission R7 \citep{Mainzer2014a, Meisner2017a}. We measured the brightness of the host using LAMBDAR\footnote{\href{https://github.com/AngusWright/LAMBDAR}{https://github.com/AngusWright/LAMBDAR}} \citep[Lambda Adaptive Multi-Band Deblending Algorithm in R;][]{Wright2016a} and the methods described in \citet{schulze21}. We also measured the host magnitudes from our UVOT templates on MJD\,=\,60681.2. The measurements are listed in Table \ref{tab:host}. 

We modeled the host galaxy SED with the software package \texttt{Prospector} version 0.3 \citep{Leja2017a}. \texttt{Prospector} uses the Flexible Stellar Population Synthesis (FSPS) code \citep{Conroy2009a} to generate the underlying physical model and \texttt{python-fsps} \citep{ForemanMackey2014a} to interface with FSPS in \texttt{python}. The FSPS code also accounts for the contribution from diffuse gas (e.g., H~{\sc ii} regions) based on the Cloudy models from \citet{Byler2017a}. Furthermore, we assumed a Chabrier initial mass function \citep{Chabrier2003a} and approximated the star formation history (SFH) by a linearly increasing SFH at early times, followed by an exponential decline at late times (functional form $t \times \exp\left(-t/\tau\right)$). The model was attenuated with the \citet{Calzetti2000a} model. Finally, we used \texttt{dynesty} to sample the posterior probability function. The resulting SED fit and the best-fit parameters are shown in Fig. \ref{fig:hostfit}.

The absolute magnitude of the host galaxy in the optical is between $-17.28\pm0.05$ (LS $g$ band) and $-17.89\pm0.11$~mag (LS $z$ band). Assuming the $g$-band value is comparable to the $B$-band absolute magnitudes reported for the \citet{kangas22} sample, the host galaxy of SN~2023gpw would be somewhat fainter than the median ($-18.7$~mag) or the average ($-17.9$~mag). The host-galaxy extinction from the fit is very minor ($E(B-V)_{\mathrm{star}} = 0.05^{+0.06}_{-0.03}$~mag), less than $2\sigma$ from zero extinction. In terms of stellar mass (log~$M/\mathrm{M}_\odot = 8.67^{+0.17}_{-0.13}$), the host is roughly average compared to other SLSN~II hosts \citep{kangas22}; the star formation rate ($0.03^{+0.05}_{-0.03}$~M$_\odot$~yr$^{-1}$), though, is rather low. The host galaxy lies on the low-mass tail of the host galaxy distribution of normal SNe~II at $z>0.08$ \citep{schulze21}. 

%--------------------------------------------------------------------
\section{Discussion}
\label{sec:disco}

\subsection{Excluding the tidal disruption event scenario}

Despite the (by eye) near-nuclear location of SN~2023gpw, we find a distance of $\sim$1.5~kpc or $\sim0\farcs8$ between the centroid of the SN and the nucleus of the host galaxy (see Sect. \ref{sec:host}), making a TDE origin for it highly unlikely. Nonetheless, as an additional check, we here compare it to TDEs. The spectral features observed in SN~2023gpw are dominated by hydrogen and helium features early on, similarly to the H+He class of TDEs \citep{arcavi14,vanvelzen21}, but several differences are apparent. The host galaxy itself is a blue, faint, star-forming spiral galaxy with a relatively weak nucleus, which is atypical for a TDE host galaxy -- these tend to be post-merger, green-valley galaxies with centrally concentrated stellar distributions \citep{hammerstein21}. TDEs generally show declining blackbody radii, roughly constant temperatures and a relation between temperature and radius, $T \propto R^{-1/2}$ \citep{vanvelzen20,vanvelzen21,hammerstein23}, which clearly contradict the observed blackbody evolution of SN~2023gpw (see Fig. \ref{fig:bbparams}). The NUV excess we see in our SEDs is also abnormal for a TDE, where typically a single blackbody captures the optical and NUV evolution \citep[e.g.][]{vanvelzen20}. We can therefore rule out a TDE origin for SN~2023gpw.

\subsection{SN~2023gpw compared to other SLSNe II}

Light-curve-wise, SN~2023gpw does not particularly stand out from other SLSNe~II in the $g$ band. It is among the more luminous objects of the class, but not extremely so; its luminosity places it close to the border of the ``IIP/L-like'' and ``08es-like'' subclasses among SLSNe~II that \citet{kangas22} divide their sample into. While SN~2023gpw is initially somewhat hotter and bluer than other SLSNe~II, this does not last very long. The host galaxy of SN~2023gpw does not stand out compared to those of other SLSNe~II; in terms of these subgroups, it is perhaps more similar to the ``IIP/L-like'' group, but the sample of ``08es-like'' SNe is small enough that the difference is not clear.

However, two things set SN~2023gpw apart. One is that in all other SLSNe~II, the UV excess is visible at all observed epochs \citep{miller09,kangas22}, while in SN~2023gpw it appears between +4 and +18~d, when the blackbody temperature declines from 15\,000 to 11\,000~K. The extra emission may simply be less clear at higher temperatures where the blackbody peak coincides with it. The more important thing is the steepening light curve in the redder bands, possibly similar to normal SNe~II where it signals the full recombination of the hydrogen and the transition to the radioactive tail. A steepening like this has not been observed in any other SLSN~II, and can indeed be excluded in some, including the prototypical SN~2008es \citep{kornpob19}. A drop of multiple magnitudes is not necessarily related to recombination. Similar features can appear if the dominant energy input mechanism turns off; for example, in interacting SNe when the forward shock reaches the outer edge of the CSM \citep{khatami24}.

Out of the models included in \texttt{MOSFiT} that we applied in this study, only the CSI model replicates the light curve, including this steepening. This is in contrast to the samples of \citet{inserra18} and \citet{kangas22}, where the magnetar model was either favored over the CSI model or was roughly equally effective when it comes to light-curve fitting, though some signs of interaction were seen, such as a UV excess over a blackbody and/or late-time multi-component H$\alpha$ profiles. Even in SN~2008es, \citet{kornpob19} could not exclude the magnetar model with a time-dependent gamma-ray trapping efficiency, though their late-time data favored the CSI model. 

Multi-component `flash' features are not seen in most SLSNe~II. In some cases, this can be simply because no early spectra exist, except possibly at very low resolution, and their existence at similar times since the explosion cannot often be excluded. In the combined \citet{inserra18} and \citet{kangas22} sample, SNe~2019zcr, 2019aanx and 2020bfe have pre-peak spectra ($-16$, $-11$ and $-14$~d respectively) with a possibly multi-component line profile of H$\alpha$ as in SN~2023gpw, but He~{\sc ii}$~\lambda$4868 is not seen in any spectrum. The earliest spectra of SN~2008es in \citet{gezari09} do show a He~{\sc ii}$~\lambda$4868 as well as H$\alpha$ at $\sim-10$~d, but neither has a narrow component.

The later photospheric spectra show some differences from other SLSNe~II as well. The spectra of SN~2008es (Fig. \ref{fig:spexcomp1}), and by extension other 2008es-like SNe \citep{kangas22}, show broader, more symmetric H$\alpha$ profiles, but similar metal (iron?) lines in the blue part of the spectrum. Meanwhile, several SLSNe~II show spectra similar to somewhat less-luminous SNe~II (Fig. \ref{fig:spexcomp2}), either normal SNe~II with P~Cygni line profiles or interacting SNe~II such as SN~1979C \citep{panagia80} and SN~1998S \citep{fassia01}. The objects without clear P~Cygni profiles in H$\alpha$ still often have other typical SN~II features, such as clearly skewed H$\alpha$ profiles with suppressed wings, Ca~{\sc ii} NIR triplet emission and/or strong Na~{\sc i} + He~{\sc i} absorption, none of which are clearly seen in SN~2023gpw. The peak luminosity of SN~2023gpw is on the borderline between these two subgroups as stated above, and its spectra show similarity to both groups but have no exact match among other SLSNe~II. 

The polarization of only one SLSN~II has been measured previously: \citet{inserra18} obtained an early polarization degree of $p = (0.94 \pm 0.17)\%$ for PS15br, implying an asymmetry of 10--15\% assuming an ellipsoidal photosphere; however, the authors note that it is difficult to distinguish whether the polarization is intrinsic to PS15br or interstellar. The polarization degree in SN~2023gpw is consistent with this, $p = (0.66 \pm 0.33)\%$ (see Sect. \ref{sec:pol}); though it is only $2\sigma$ from zero. Similar pre-peak numbers have also been seen in SLSNe~I \citep{pursiainen23}. As a comparison, \citet{leonard00} found a pre-peak 2\% polarization in SN~1998S, which was implied to have a disk-like CSM because of the polarization and multi-peaked line profiles. \citet{nagao24}, on the other hand, found polarization degrees of $\lesssim$1.2\% in the photospheric phases of normal SNe~II; often the polarization increases in the tail phase, revealing asymmetries in the inner ejecta. In a similar vein, the late-time H$\alpha$ profiles of PS15br and SN~2013hx had multiple components, implying asymmetric CSM \citep{inserra18}. We cannot say whether either kind of evolution occurs in SN~2023gpw.

\subsection{Asymmetry and evolution of the photosphere}

In the spectra, as discussed in Sect. \ref{sec:specs}, we see `flash' emission lines of the Balmer series and He~{\sc ii}~$\lambda4686$. At the epoch of our polarization measurement, these lines are still visible on top of an otherwise featureless continuum. Therefore, the photosphere is -- at least partially -- within the CSM and does not necessarily reflect the shape of the ejecta. The CSM seems to be only slightly aspherical, but unfortunately, this too is difficult to say for certain, as it depends on the angle of the ISP and the viewing angle. However, there is another sign of asymmetry in the ejecta. The velocity of the H$\alpha$ emission and He~{\sc i}~$\lambda5876$ absorption in the ejecta is lower than that of the photosphere from the blackbody fits by about a factor of 2. This suggests an aspherical, possibly bipolar ejecta, in which case an observer at some viewing angles could see lower velocities in the radial than the transverse direction (lines vs. photosphere expansion).

The expansion of the photosphere, according to our blackbody fits, does not appreciably decelerate, and is instead fairly constant at $\sim$9100~km~s$^{-1}$ until $\gtrsim$55~d after the peak, indicating that the photosphere does not recede below an outer ejecta layer for a long time. In our line of sight, this layer corresponds to a much lower velocity ($\sim$4500~km~s$^{-1}$) because of the asymmetry of the ejecta. For the photosphere to stay at a constant velocity throughout our observations and even before the peak, the ejecta requires a relatively sharp outer edge -- this could also result in a weak P~Cygni absorption in H$\alpha$ \citep{schlegel96}. Analogously to a normal SN~II, the opacity of the ejecta can drop dramatically after cooling below the recombination temperature, which is almost reached in our blackbody fits; this can cause a drop in the light curve as the photosphere finally recedes.

\subsection{CSM interaction}

The flash features in SN~2023gpw are in broad strokes similar to those seen in normal SNe~II \citep[e.g.][]{bruch21,bruch23}, as they consist of a narrow component with broad wings attributed to electron scattering, which weaken over time and disappear between the peak and +12~d, when the He~{\sc ii}~$\lambda4686$ seems to have been replaced by a P~Cygni profile of He~{\sc i}~$\lambda5876$. Thus, with our constraints on the explosion date (within a few days of the discovery according to \texttt{MOSFiT}), we can say that these lines disappear roughly between 32 and 45 days after the explosion. Assuming the disappearance of the lines is because the SN ejecta catches up to and sweeps up the CSM where they originate, the ejecta velocity ($\sim$9100~km~s$^{-1}$) we estimate based on the expansion of the blackbody radius (see Sect. \ref{sec:BB}) leads to an extent of between $2.5 \times 10^{15}$ and $3.5 \times 10^{15}$~cm for this `confined' CSM. We note, though, that the radius of the blackbody was calculated assuming a spherical photosphere, and thus the velocity can be considered an average over all solid angles. Along the line of sight, the velocity is smaller by about a factor of 2 and thus the extent of the CSM can be closer to $\sim$1.5 $\times 10^{15}$~cm. In many SNe~II, flash features tend to disappear in a few to several days \citep{khazov16,bruch21,wynn24} implying an extent of $< 10^{15}$~cm for the confined CSM assuming a representative ejecta velocity of 10\,000~km~s$^{-1}$; though we note that the recent, well-observed SN~2023ixf exhibited narrow lines until $\sim$15~d \citep{zimmerman24}. In any case, such features last relatively long in SN~2023gpw, and the CSM responsible for them seems more extended than in normal SNe~II.

The disappearance of the flash features does not indicate the end of all CSI. After this time, the lack of narrow lines could in principle be due to any strong interaction being hidden behind the ejecta photosphere if, for example, the ejecta overruns an aspherical, possibly disk-like CSM \citep[e.g.][]{smith15}. In this case, the CSI would keep contributing to the luminosity without narrow lines. An aspherical interaction region within the ejecta should result in an aspherical photosphere, as our velocity discrepancy indicates \citep[e.g.][]{smith17}. Again, the ISP and the viewing angle have an effect on the resulting polarization. Alternatively, an optically-thick CSM with a sharp outer edge could explain the lack of narrow lines \citep[][]{moriya12}. In this case, only a small part of the CSM near the outer edge is optically thin. Narrow lines in the spectrum would only appear in this scenario while the shock is in the optically thin outer part of the CSM, leading to events apparently without strong narrow lines if this part is thin. In the case of SN~2023gpw, we could see narrow lines with electron-scattering wings from the outer layers of this CSM until they are shocked by the ejecta.

One signature of CSI we see after the flash features disappear is the UV excess in the form of a flat NUV spectrum (see Sect. \ref{sec:BB}). A UV excess is also seen in all SLSNe~II with UV data shown in \citet{kangas22}, and in some other, less luminous SNe~II such as SN~1979C \citep{panagia80} and SN~2023ixf \citep{bostroem24}.\footnote{SN~2020yue does not show such an excess, but this object was later re-classified a TDE by \citet[][]{yao23}.} In SN~2023gpw, this excess appears a few weeks after the peak -- about the same time as the broad emission lines. The epoch when the excess appears was missed in the objects shown by \citet{kangas22}, but in SN~2008es, the excess predates the broad lines \citep{miller09}. To create such an excess, one needs a source of UV photons that aren't reprocessed in the ejecta and therefore originate outside the photosphere, making CSI an obvious candidate. If UV photons can escape, a large portion of CSI energy is indeed expected to emerge in the UV \citep[e.g.,][]{fransson84,dessart22}. The excess could, however, be created through weaker interaction with a more tenuous outer CSM, to which the UV is more sensitive than the optical \citep{bostroem24}. In fact, in some strongly interacting objects, such as SN 2013L \citep{taddia20}, a clear UV excess is not seen, likely because the UV photons are thermalized while the forward shock is still in the dense CSM. 

We therefore propose that, early on, before and around the peak of the light curve, CSI occurs between the ejecta and a dense, slightly asymmetric, CSM confined to within a~few~$\times10^{15}$~cm. An extended envelope is also a possibility, but this radius ($\gtrsim15000$~R$_\odot$) surpasses the radii of quiescent supergiant stars by an order of magnitude \citep{levesque17}. The peak of the light curve in this scenario is expected to nearly coincide with the emergence of the shock from within the optically-thick part of the CSM \citep{moriya12}, which is consistent with our observations. Further out, there is a more tenuous CSM, and further interaction with it creates the UV excess. 

If interaction with a dense CSM or extended envelope occurs during the rise, the early photosphere should exhibit at least the radius of the optically thick part of the CSM/envelope and an increasing temperature. The earliest blackbody fits are indeed consistent with this expectation, but not very robust as they are based on only two or three optical bands. Any significant deceleration of the expansion of the photosphere (necessary if bulk deceleration of the ejecta is significant) would result in the constant-velocity line (Fig. \ref{fig:bbparams}) intersecting zero radius before the explosion, which is not the case. The implication is that despite the interaction, significant deceleration does not occur in the bulk of the ejecta, and therefore that the CSM mass is small compared to that of the ejecta -- though deceleration can happen in outer layers initially faster than $\sim9000$~km~s$^{-1}$.

We can roughly estimate the mass-loss rate from the peak pseudo-bolometric luminosity $L_\mathrm{peak}$ of the SN using the following relation \citep[e.g.][]{smith17}, assuming the luminosity at this point is interaction-dominated:
\begin{equation}
    \frac{dM}{dt} = \frac{2L_\mathrm{peak}}{\epsilon} \frac{v_{\mathrm{CSM}}}{v^3_{\mathrm{ej}}}~,
\end{equation}
where $v_{\mathrm{CSM}}$ is the CSM velocity and $v_{\mathrm{ej}}$ the ejecta velocity. We note that the luminosity is pseudo-bolometric, which underestimates the mass loss, but at the same time the assumption of being interaction-dominated may overestimate it. We first assume that the width of the narrower emission component of H$\alpha$, intrinsically roughly 700~km~s$^{-1}$, corresponds to the velocity of material ejected previously (i.e., $v_{\mathrm{CSM}} = 700$~km~s$^{-1}$). With $L_\mathrm{peak}\sim2.8\times10^{44}$~erg~s$^{-1}$ and $\epsilon = 0.5$, we obtain $\frac{dM}{dt} \sim 1.9\, \mathrm{M}_\odot \,\mathrm{yr}^{-1}$. This rate is somewhat extreme, but perhaps plausible in an LBV giant eruption \citep{smith14}. With the radius of the confined CSM (a few $\times 10^{15}$~cm) we obtain from the timing of the flash features, the mass loss responsible for that CSM would have happened on the order of a year before the SN explosion. If, however, this component is created through electron scattering in unshocked CSM, we can instead assume a velocity more typical for the wind of a luminous blue variable (LBV), $v_{\mathrm{CSM}} = 100$~km~s$^{-1}$ \citep{smith17}, which is also closer to the narrow lines we see at +29 and +43~d. Doing so, we obtain $\frac{dM}{dt} \sim 0.27\, \mathrm{M}_\odot \,\mathrm{yr}^{-1}$ and a timing within several years of the SN.

\subsection{The power source and progenitor of SN~2023gpw}

As stated above, the \texttt{MOSFiT} magnetar or (continuous) fallback models seem to fail to reproduce the light curve of SN~2023gpw and the steepening in the light curve (see Sect. \ref{sec:mosfit}). The obvious conclusion of this would be to favor the CSI model, which can replicate the observed light curve especially with $s=0$. However, in the CSI scenario, the energy source is ultimately the kinetic energy of the ejecta, regulated by the efficiency of the conversion of this energy into radiation, $\epsilon$ \citep[as an example, a range of 0.3 to 0.7 for $\epsilon$ is seen in simulations by][with higher CSM-to-ejecta mass ratios resulting in more efficiency]{dessart15}. In SN~2023gpw, similarly to the few most luminous objects in \citet{kangas22}, the total radiated energy is on the order of $10^{51}$~erg. This may represent a problem for the neutrino-driven CCSN mechanism, as neutrino-driven simulations are unable to exceed $\sim2\times10^{51}$~erg \citep{janka12}. If it does, another power source, such as a central engine, may also be required. 

In our case, a neutrino-driven SN would require on the order of $\gtrsim45\,\%$ of the total kinetic energy to be converted to reproduce the radiated energy of $\gtrsim8.9 \times 10^{50}$~erg we estimate (see Sect. \ref{sec:LC}), which is, in itself, not necessarily a problem in the CSI model, but would cause significant deceleration in the ejecta (by a few thousand km~s$^{-1}$). It is possible that, if the CSM is very asymmetric, such deceleration does occur along the line of sight while most of the ejecta expands unimpeded. This would, however, mean that a large fraction of the ejecta does not take part in the interaction, and that the kinetic energy again becomes a problem; a neutrino-driven explosion would require at least 60\,\% of the ejecta to decelerate to 5000~km~s$^{-1}$. It is more likely that the ejecta itself is asymmetric to begin with.

We next attempt to estimate an ejecta mass for SN~2023gpw based on the rise time of the light curve using the equation
\begin{equation}
 \tau_m = \left(\frac{2\kappa M_\mathrm{ej}}{\beta cv} \right)^{1/2}~,
\end{equation}
where $\tau_m$ is the diffusion timescale, $\beta \approx 13.8$ is an integration constant, and $v$ is a scaling velocity for homologous expansion \citep{arnett80,arnett82}. We fix $\kappa$ at 0.34~cm$^2$~g$^{-1}$ as before. We approximate the diffusion timescale as the rise time, $\sim$30~d, and the scaling velocity as the expansion velocity of the blackbody photosphere. The resulting ejecta mass is $\sim$3~M$_\odot$. With this mass and the blackbody expansion velocity, the estimated kinetic energy would be $E_k \approx 0.3 M_{ej} v_{ej}^{2} \approx 1.5\times10^{51}$~erg. Here the radiated energy would be on the order of $\gtrsim60\,\%$ of the kinetic energy, which is again problematic. 

This is an indication that the Arnett formula, which assumes a central power source and diffusion through the whole ejecta, is inapplicable to SN~2023gpw, as the diffusion only occurs outwards from the interaction region, and the true ejecta mass is likely much larger. If only a small fraction of the total kinetic energy is emitted, it must be at least several~$\times10^{51}$~erg, which, at a velocity of 9100~km~s$^{-1}$, translates to a mass at least on the order of 10~M$_\odot$. This rough lower limit is compatible with our fitting results with the \texttt{csm} model. However, we note that the CSM mass in the $s=0$ fit is large compared to the ejecta mass, which is unlikely if the CSM is symmetric but a large fraction of the kinetic energy is not emitted. Furthermore, fits using the \texttt{csm} model cannot account for any asymmetry. As an order-of-magnitude estimate, the ejecta mass in the $s=2$ fit is $30^{+3}_{-2}$~M$_\odot$, resulting in $E_k \sim 1.4 \times 10^{52}$~erg. In this case, the fraction of total kinetic energy converted to radiation would only be $\sim$0.05. 

If energies in excess of a neutrino-driven SN are required, we must consider the magnetar spin-down model. At first glance, a magnetar power source is disfavored by the \texttt{MOSFiT} modeling (Sect. \ref{sec:mosfit}) because of the drop in the red bands of the light curve. The generalized magnetar model of \citet{omandsarin24} similarly cannot reproduce this drop. Magnetar models specifically constructed for hydrogen-rich SLSNe, however, can produce a plateau and a drop analogous to normal SNe~II with some parameters \citep{orellana18}. One can alternatively obtain intermediate cases with a short plateau phase and a drop in the (bolometric) light curve, or a smoother light curve with no drop. Notably, some of these models also recreate the observed constant photospheric velocity, as most of the ejecta is pushed from within into a dense shell that it takes a long time for the photosphere to recede through. 

We note that other simulations do not necessarily agree. In models of magnetar-powered SLSNe~II by \citet{dessart18}, the same qualitative light-curve shapes with plateaus can be generated, but while the influence of a magnetar does slow down the photospheric velocity decline, it is too steep for SN~2023gpw nonetheless. In three-dimensional magnetar-driven SN simulations by \citet{suzuki19}, the ejecta is also being pushed from the inside, but the hot, fast magnetar wind breaks through it, resulting in a thick, clumpy dense ejecta instead of a thin shell, with a relatively shallow outer density profile. This could explain the UV excess, but seems incompatible with the constant photospheric velocity. However, \citet{suzuki19} do not include radiative transfer in their simulation and note that doing this might suppress the breakout.

The decline of the pseudo-bolometric light curve of SN~2023gpw only slows down slightly before the solar conjunction, with no plateau, but, during the gap, presumably steepens as the red bands do. This is still qualitatively similar to the intermediate models of \citet{orellana18}. Another problem, however, is that the aforementioned models have no early CSI. SN~2023gpw is more luminous at peak than any model shown by \citet{orellana18} or \citet{dessart18}, by a factor of a few, and declines faster, which could be attributed to CSI analogously to a normal SN~II \citep{morozova17}. We also note that the models discussed here form a sparse grid, with the ejecta mass fixed at 10~M$_\odot$ in the \citet{orellana18} and \citet{suzuki19} simulations and either 11.9 or 15.6~M$_\odot$ in \citet{dessart18}. It is possible that by adjusting the parameters and taking into account the early CSI, the observed properties could be reproduced. We encourage modeling this scenario in further detail, including multi-band light curves, to determine whether it is applicable. We also note that these simulations assume spherical symmetry for the ejecta and the spin-down energy input, which seems to be disfavored by our observations.

If the ejecta mass in SN~2023gpw can be on the order of $30$~M$_\odot$, there is a possibility of an extremely massive progenitor star, and a pulsational pair-instability SN \citep[PPISN;][]{woosley17} scenario should be discussed. A star with a helium core of 35--65~M$_\odot$ can experience pair instability that does not fully unbind the star, but instead proceeds in pulses (hence pulsational pair instability, PPI) that eject large amounts of mass. Transients from such objects can include the mass ejections themselves, collisions between CSM shells, or the final SN at the end of such a star's life -- if one can happen -- plus CSI. However, \citet{woosley17} states that a SLSN is difficult to produce with PPI. The radiated energy from PPISN models is not expected to exceed $5\times10^{50}$~erg, and shell velocities should be much lower than the blackbody velocity observed in SN~2023gpw. 

After PPI, the core is unlikely to explode as a SN, but it might do so in a collapsar-like scenario \citep[see also][]{woosley93}. \citet{woosley17} brings up a situation where after PPI, the core partially collapses into a black hole and accretion onto it powers a bipolar outflow, which, similarly to the magnetar central engine, can then interact with the previously ejected mass and artificially condense the ejecta from the inside. The accretion could produce a kinetic energy in excess of $10^{52}$~erg, as is often seen in broad-lined SNe~Ic (SNe~Ic-BL) accompanying gamma-ray bursts (GRBs), also the products of collapsars \citep[e.g.][]{woosley99}. Other SNe~Ic-BL can also exhibit similar energies even without a GRB \citep[e.g.][]{taddia19}, if a relativistic jet does not break out of the ejecta. Whether the progenitor star experienced PPI or not, a black hole as the central engine may explain SN~2023gpw. The \texttt{MOSFiT} fallback model suffers from the same late-time problem as the magnetar model, but it assumes continuous accretion. If, instead, accretion turns off after some time, a steepening decline can be accommodated.

%--------------------------------------------------------------------
\section{Conclusions}
\label{sec:concl}

We have presented photometric and spectroscopic follow-up observations and analysis of SN~2023gpw, a SLSN~II with early `flash' emission lines and later dominated by broad spectral lines. We have performed blackbody and light-curve modeling and compared SN~2023gpw to other SLSNe~II. Our conclusions are as follows:

\begin{itemize}
    \item With a rise time of $t_\mathrm{rise,1/e} = 16.6^{+3.2}_{-3.0}$~d and a $g$-band peak absolute magnitude of $M_{g,\mathrm{peak}} = -21.58\pm0.03$~mag, SN~2023gpw is a relatively fast-rising SLSN~II, but neither its luminosity nor its rise time are unprecedented among SLSNe~II. Its host galaxy is fainter and less massive than most SN~II hosts, but normal for a SLSN~II host.
    \item The UV photometry exhibits an excess over a blackbody, which is so far ubiquitous among SLSNe~II and also seen in some normal SNe~II. SN~2023gpw is, however, so far the only SLSN~II where its appearance, coinciding with that of the broad spectral lines, is documented.
    \item The blackbody radius, when ignoring the bluest UV data after the peak, expands at a constant velocity of $\sim9100$~km~s$^{-1}$. Extrapolation of this velocity to earlier epochs results in zero radius close to the explosion date ($\sim-30$~d), indicating no significant bulk deceleration in the ejecta throughout the observed evolution. The photosphere does not seem to recede past the outer ejecta until $\gtrsim+50$~d.
    \item The light curve indicates an abrupt drop of several magnitudes in the red bands during a solar conjunction, between $\sim+80$ and $\sim+180$~d, after which the SN is no longer detected. This is hitherto unprecedented among SLSNe~II, and may correspond to the epoch when the photosphere finally recedes into lower-velocity ejecta.
    \item The flash features, their disappearance around the light-curve peak and the later UV excess indicate interaction with a dense, confined CSM early on, followed by ongoing weaker interaction with a more tenuous CSM. 
    \item Likely polarization at $p=(0.66\pm0.33)\%$ is observed during the early interaction, suggesting a slightly aspherical CSM and/or an aspherical ejecta interacting with it. The discrepancy in expansion velocities measured from the blackbody fits and from the spectral lines also indicates asymmetry. 
    \item The total radiation emitted by SN~2023gpw is estimated as $\gtrsim8.9\times10^{50}$~erg, which, in a normal neutrino-driven SN, would require a large fraction of the total kinetic energy to be converted to emission. However, due to the lack of bulk deceleration, the observed interaction is unlikely to result in such an efficient conversion. Instead, it is likely that the kinetic energy is on the order of $10^{52}$~erg, and the ejecta mass therefore $\gtrsim10$M$_\odot$.
    \item The light-curve drop is not reproduced by the magnetar model in \texttt{MOSFiT}. The light curve and constant photospheric velocity are, however, qualitatively similar to some models of magnetar-powered hydrogen-rich SLSNe \citep{orellana18}, but more modeling work is required to determine if this scenario, combined with early interaction, can reproduce the full light curve.
    \item Another scenario for producing the required kinetic energy and ejecta mass is one where the core of a very massive progenitor star, possibly after PPI episodes that produce the CSM, collapses into a black hole. Fall-back accretion onto this black hole then results in a bipolar outflow heating the ejecta from the inside, until the accretion stops, causing the drop in brightness. 
\end{itemize}

\section*{Data availability}

All spectra presented in this paper are available for download on WISEREP \citep{yaron12} at \url{https://www.wiserep.org/object/23362}.
\\

\begin{acknowledgements}

We thank the anonymous referee for useful comments, and Georgios Dimitriadis for re-reducing the INT spectrum. \\

T.K. acknowledges support from the Research Council of Finland project 360274. R.K. and T.L.K. acknowledge support from the Research Council of Finland (340613). \\

M.S. is funded by the Independent Research Fund Denmark (IRFD) via Project 2 grant 10.46540/2032-00022B and by the Aarhus University Nova grant\# AUFF-E-2023-9-28. \\

N.E.R. acknowledges support from the PRIN-INAF 2022, `Shedding light on the nature of gap transients: from the observations to the models' and from the Spanish Ministerio de Ciencia e Innovaci\'on (MCIN) and the Agencia Estatal de Investigaci\'on (AEI) 10.13039/501100011033 under the program Unidad de Excelencia Mar\'ia de Maeztu CEX2020-001058-M. \\

C.P.G. acknowledges financial support from the Secretary of Universities and Research (Government of Catalonia) and by the Horizon 2020 Research and Innovation Programme of the European Union under the Marie Sk\l{}odowska-Curie and the Beatriu de Pin\'os 2021 BP 00168 programme, from the Spanish Ministerio de Ciencia e Innovaci\'on (MCIN) and the Agencia Estatal de Investigaci\'on (AEI) 10.13039/501100011033 under the PID2023-151307NB-I00 SNNEXT project, from Centro Superior de Investigaciones Cient\'ificas (CSIC) under the PIE project 20215AT016 and the program Unidad de Excelencia Mar\'ia de Maeztu CEX2020-001058-M, and from the Departament de Recerca i Universitats de la Generalitat de Catalunya through the 2021-SGR-01270 grant.\\

T.L.K. acknowledges a Warwick Astrophysics prize post-doctoral fellowship made possible thanks to a generous philanthropic donation. \\

T.E.M.B. is funded by Horizon Europe ERC grant no. 101125877. \\

The Starlink software \citep{currie14} is currently supported by the East Asian Observatory. \\

This study is based on observations made with the Gran Telescopio Canarias (GTC), installed in the Spanish Observatorio del Roque de los Muchachos of the Instituto de Astrofísica de Canarias, in the island of La Palma. \\ 

Access to the Las Cumbres Observatory was made possible via an allocation by OPTICON (program 22A/012, PI Stritzinger). This project has received funding from the European Union's Horizon 2020 research and innovation programme under grant agreement No 101004719. \\

Based on observations obtained with the Samuel Oschin Telescope 48-inch and the 60-inch Telescope at the Palomar Observatory as part of the Zwicky Transient Facility project. ZTF is supported by the National Science Foundation under Grant No. AST-2034437 and a collaboration including Caltech, IPAC, the Weizmann Institute of Science, the Oskar Klein Center at Stockholm University, the University of Maryland, Deutsches Elektronen-Synchrotron and Humboldt University, the TANGO Consortium of Taiwan, the University of Wisconsin at Milwaukee, Trinity College Dublin, Lawrence Livermore National Laboratories, IN2P3, University of Warwick, Ruhr University Bochum, Cornell University, and Northwestern University. Operations are conducted by COO, IPAC, and UW. \\

\end{acknowledgements}

\bibliographystyle{aa}
\bibliography{kangas_23gpw_20250923}
\clearpage

\begin{appendix}

\section{Data reduction}
\label{sec:redu}

The ALFOSC images were reduced using the custom pipeline \texttt{Foscgui},\footnote{\texttt{Foscgui} is a graphic user interface aimed at extracting SN spectroscopy and photometry obtained with FOSC-like instruments. It was developed by E. Cappellaro. A package description can be found at \url{http://sngroup.oapd.inaf.it/foscgui.html}.} which performs bias subtraction and flat-field correction. The \texttt{IRAF} package \texttt{notcam}\footnote{\url{https://www.not.iac.es/instruments/notcam/quicklook.README}} was used to perform sky subtraction, flat-field correction, distortion correction, bad-pixel masking and co-addition of the NOTCam NIR images. The ZTF data reduction and pipelines are managed by the Infrared Processing and Analysis Center (IPAC) at Caltech as described by \citet{masci19}. The SEDM imaging data were processed using point-spread-function (PSF) photometry, and the magnitudes were calibrated against either Sloan Digital Sky Survey (SDSS) or Panoramic Survey Telescope and Rapid Response System 1 \citep[Pan-STARRS1;][]{flewelling20} reference images, using \texttt{Fpipe} \citep{Fremling2016}. LT data were processed by a similar custom-built software \citep{Taggart2020}. ZTF and LT $griz$ magnitudes were calibrated to the Pan-STARRS1 photometric system and $u$-band data to the SDSS system.

Aperture photometry was performed on the LCOGT and NOT images using Starlink\footnote{\url{http://starlink.eao.hawaii.edu/starlink}} \texttt{Gaia}\footnote{\url{http://star-www.dur.ac.uk/~pdraper/gaia/gaia.html}} \citep{currie14}. The zero points of each image were calibrated using field stars. In the $gri$ bands, the AAVSO Photometric All-sky Survey \citep[APASS\footnote{\url{https://www.aavso.org/apass}};][]{henden14} was used for calibration. In the $u$ band, which is not included in APASS, the field stars were calibrated using a photometric standard field included in SDSS DR 17 \citep{sdss-dr17} and observed in photometric conditions. $JHKs$-band magnitudes were calibrated using stars included in the Two Micron All Sky Survey \citep[2MASS;][]{skrutskie06} catalog.

For NUV photometry, we used the tasks \texttt{uvotimsum} and \texttt{uvotsource} in \texttt{HEASOFT}\footnote{\url{https://heasarc.gsfc.nasa.gov/}} v.6.31.1 to co-add individual exposures and measure the source brightness, respectively. A circular aperture with a 5-arcsecond radius was used to extract the source flux, with background flux subtracted using a 10-arcsecond-radius circular region. The same regions were used to extract the host galaxy flux from the late-time template images; this flux was then subtracted from the other NUV epochs.

For the optical imaging performed after MJD\,=\,60289, we used the \texttt{IRAF} tasks \texttt{geomap} and \texttt{geotran} to align our images to pre-explosion template images downloaded from the Pan-STARRS1 Image Cutout Service\footnote{\url{http://ps1images.stsci.edu/cgi-bin/ps1cutouts}}. After this, the \texttt{ISIS 2.2} package based on the Optimal Image Subtraction method \citep{alard98,alard00} was used to scale the templates to our images and subtract the host galaxy contribution. 

Polarimetry data were reduced with a custom pipeline that uses \texttt{photutils} \citep{bradley22} for the photometry. The optimal aperture size was chosen to be 2 times the FWHM (i.e. $2 \times 5.0 = 10.0$ pixels) in order to enclose the majority of the flux in the aperture and avoid inducing spurious polarisation due to the different PSF elongation of the sources in the ordinary and extraordinary beams, respectively, that is occurring in the imaging polarimetry mode of ALFOSC \citep[see][and discussions therein]{leloudas17,pursiainen23}. The resulting mean S/N of our observations was $\sim$ 288 in $V$ and 273 in $R$. 

NOT/ALFOSC and GTC spectra were reduced -- bias- and flat-corrected, wavelength- and flux-calibrated, and corrected for telluric absorption -- using \texttt{Foscgui}. The INT/IDS spectra were reduced through similar steps using standard \texttt{IRAF} and \texttt{PyRAF}\footnote{\url{https://github.com/iraf-community/pyraf}} routines, with the final flux calibration and telluric line removal done with \texttt{IDL} routines as in \citet{silverman12}. The two INT/IDS spectra were rather noisy due to the proximity of the moon, and being less than 1 day apart, were normalized and average-combined into one spectrum at an exposure-weighted average epoch of +11.6~days.

Reductions of P200/DBSP spectra were carried out using \texttt{DBSP\_DRP} \citep{roberson21}, based on \texttt{PypeIt} \citep{prochaska20}. The DBSP spectra were taken close to solar conjunction, when the SN was setting early, and were therefore taken during twilight, resulting in difficulties with removing the telluric absorption. The final spectra (Sect. \ref{sec:specs}) show strong absorption lines that are most likely a result of this. SEDM spectra were processed by the Python-based \texttt{pysedm} pipeline \citep{rigault19,kim22}.  The three last SEDM spectra, taken between +31 and +34~d, were median-combined with an average epoch of +33~d.

\section{Supplementary tables}
\label{sec:tables}

The photometric measurements listed below are not corrected for Galactic extinction. $K$-corrections have not been applied either. When host-galaxy subtraction has been applied, it is explicitly noted in the appropriate table.

\FloatBarrier

\begin{table}[h!]
\caption{NOT/NOTCam photometry of SN~2023gpw. }
\label{tab:notcam}
\begin{tabular}{cccc}
\hline
MJD & Filter & Magnitude & Error \\
 & & (mag, Vega) & (mag, Vega) \\
\hline
60109.9 & $J$ & 16.40 & 0.05 \\
60109.9 & $H$ & 16.42 & 0.05 \\
60109.9 & $Ks$ & 16.23 & 0.09 \\
\hline
\end{tabular}
\end{table}

\begin{table}[h!]
\caption{NOT/ALFOSC photometry of SN~2023gpw.}
\label{tab:alfosc}
\begin{tabular}{cccc}
\hline
MJD & Filter & Magnitude & Error \\
 & & (mag, AB) & (mag, AB) \\
\hline
60077.0 & $u$ & 16.69 & 0.02 \\
60077.0 & $g$ & 16.52 & 0.01 \\
60077.0 & $r$ & 16.77 & 0.03 \\
60077.0 & $i$ & 16.97 & 0.02 \\
60289.2*	& $g$ & $>$23.0 & - \\
60289.2*	& $r$ & $>$23.2 & - \\
60289.2*	& $i$ & $>$22.5 & - \\
60360.2*	& $g$ & $>$23.7 & - \\
60360.2*	& $r$ & $>$22.2 & - \\
60360.2*	& $i$ & $>$22.5 & - \\
60384.2*	& $r$ & $>$21.1 & - \\
60384.2*	& $i$ & $>$21.5 & - \\
60401.0*	& $r$ & $>$23.5 & - \\
60401.0*	& $i$ & $>$23.2 & - \\
\hline
\end{tabular}
\tablefoot{Host subtraction has been performed at late-time epochs marked with an asterisk (*).}
\end{table}

\begin{table}
\caption{LCOGT photometry of SN~2023gpw.}
\label{tab:lcogt}
\begin{tabular}{cccc}
\hline
MJD & Filter & Magnitude & Error \\
 & & (mag, AB) & (mag, AB) \\
\hline
60076.4 & $u$ & 16.56 & 0.04 \\
60076.4 & $g$ & 16.57 & 0.02 \\
60076.4 & $r$ & 16.84 & 0.02 \\
60076.4 & $i$ & 16.91 & 0.03 \\
60079.2 & $u$ & 16.35 & 0.01 \\
60079.2 & $g$ & 16.40 & 0.01 \\
60079.2 & $r$ & 16.68 & 0.03 \\
60079.2 & $i$ & 16.81 & 0.03 \\
60082.1 & $u$ & 16.30 & 0.02 \\
60082.1 & $g$ & 16.36 & 0.02 \\
60082.1 & $r$ & 16.60 & 0.03 \\
60082.1 & $i$ & 16.67 & 0.04 \\
60085.1 & $u$ & 16.33 & 0.02 \\
60085.1 & $g$ & 16.33 & 0.01 \\
60085.1 & $r$ & 16.55 & 0.03 \\
60085.1 & $i$ & 16.66 & 0.02 \\
60088.1 & $u$ & 16.38 & 0.03 \\
60088.1 & $g$ & 16.36 & 0.01 \\
60088.1 & $r$ & 16.59 & 0.02 \\
60088.1 & $i$ & 16.64 & 0.03 \\
60091.1 & $u$ & 16.46 & 0.02 \\
60091.1 & $g$ & 16.43 & 0.02 \\
60091.1 & $r$ & 16.59 & 0.01 \\
60091.1 & $i$ & 16.64 & 0.03 \\
60100.0 & $u$ & 16.86 & 0.02 \\
60100.0 & $g$ & 16.61 & 0.02 \\
60100.0 & $r$ & 16.70 & 0.03 \\
60100.0 & $i$ & 16.77 & 0.03 \\
60104.0 & $u$ & 17.05 & 0.02 \\
60104.0 & $g$ & 16.75 & 0.01 \\
60104.0 & $r$ & 16.82 & 0.03 \\
60104.0 & $i$ & 16.83 & 0.03 \\
60107.9 & $u$ & 17.20 & 0.05 \\
60107.9 & $g$ & 16.91 & 0.02 \\
60107.9 & $r$ & 16.89 & 0.02 \\
60107.9 & $i$ & 16.86 & 0.03 \\
60111.8 & $u$ & 17.59 & 0.03 \\
60115.9 & $u$ & 17.79 & 0.03 \\
60115.9 & $g$ & 17.16 & 0.02 \\
60115.9 & $r$ & 17.06 & 0.02 \\
60115.9 & $i$ & 17.00 & 0.04 \\
60118.7 & $u$ & 17.88 & 0.05 \\
60118.7 & $g$ & 17.22 & 0.02 \\
60118.7 & $r$ & 17.11 & 0.02 \\
60118.7 & $i$ & 16.99 & 0.03 \\
60127.8 & $u$ & 18.29 & 0.14 \\
60127.8 & $g$ & 17.59 & 0.07 \\
60127.8 & $r$ & 17.25 & 0.05 \\
60127.8 & $i$ & 17.09 & 0.05 \\
60133.0 & $u$ & 18.81 & 0.18 \\
60133.0 & $g$ & 17.74 & 0.08 \\
60133.0 & $r$ & 17.26 & 0.09 \\
60133.0 & $i$ & 17.28 & 0.14 \\
60138.0 & $u$ & 19.04 & 0.16 \\
60138.0 & $g$ & 17.91 & 0.09 \\
60138.0 & $r$ & 17.38 & 0.11 \\
60138.0 & $i$ & 17.58 & 0.16 \\
\hline
\end{tabular}
\end{table}

\setcounter{table}{2}

\begin{table}
\caption{LCOGT photometry of SN~2023gpw (continued).}
\label{tab:lcogt}
\begin{tabular}{cccc}
\hline
MJD & Filter & Magnitude & Error \\
 & & (mag, AB) & (mag, AB) \\
\hline
60143.0 & $u$ & 19.27 & 0.07 \\
60143.0 & $g$ & 17.97 & 0.03 \\
60143.0 & $r$ & 17.44 & 0.03 \\
60143.0 & $i$ & 17.25 & 0.03 \\
\hline
\end{tabular}
\end{table}

\begin{table}[h!]
\caption{LT/IO:O photometry of SN~2023gpw.}
\label{tab:ioo}
\begin{tabular}{cccc}
\hline
MJD & Filter & Magnitude & Error \\
 & & (mag, AB) & (mag, AB) \\
\hline
60072.9 & $g$ & 16.78 & 0.02 \\
60072.9 & $r$ & 17.04 & 0.06 \\
60072.9 & $i$ & 17.23 & 0.05 \\
60073.9 & $g$ & 16.71 & 0.03 \\
60073.9 & $r$ & 16.98 & 0.08 \\
60073.9 & $i$ & 17.19 & 0.06 \\
60082.9 & $g$ & 16.35 & 0.02 \\
60082.9 & $r$ & 16.59 & 0.07 \\
60082.9 & $i$ & 16.83 & 0.11 \\
60084.9 & $g$ & 16.37 & 0.14 \\
60084.9 & $r$ & 16.56 & 0.08 \\
60084.9 & $i$ & 16.77 & 0.07 \\
\hline
\end{tabular}
\end{table}

\begin{table}[h!]
\caption{SEDM photometry of SN~2023gpw.}
\label{tab:sedmphot}
\begin{tabular}{cccc}
\hline
MJD & Filter & Magnitude & Error \\
 & & (mag, AB) & (mag, AB) \\
\hline
60071.4 & $r$ & 17.14 & 0.08 \\
60073.2 & $r$ & 16.99 & 0.06 \\
60075.4 & $r$ & 16.85 & 0.07 \\
60079.4 & $r$ & 16.76 & 0.10 \\
60081.4 & $r$ & 16.88 & 0.15 \\
60084.3 & $r$ & 16.59 & 0.05 \\
60084.3 & $i$ & 16.51 & 0.04 \\
60098.3 & $r$ & 16.67 & 0.10 \\
60119.2 & $r$ & 17.08 & 0.05 \\
60119.2 & $g$ & 17.13 & 0.16 \\
60119.2 & $i$ & 17.04 & 0.14 \\
60120.3 & $r$ & 17.08 & 0.06 \\
60121.2 & $r$ & 17.20 & 0.18 \\
\hline
\end{tabular}
\end{table}

\begin{table}[h!]
\caption{P48 photometry of SN~2023gpw. }
\label{tab:ztfphot}
\begin{tabular}{cccc}
\hline
MJD & Filter & Magnitude & Error \\
 & & (mag, AB) & (mag, AB) \\
\hline
60049.3 & $r$ & $>$20.3 & - \\
60050.3 & $i$ & $>$19.8 & - \\
60051.3 & $g$ & $>$20.4 & - \\
60051.3 & $r$ & 20.78 & 0.32\\
60053.4 & $g$ & 20.26 & 0.28\\
60055.3 & $r$ & 19.85 & 0.22\\
60055.3 & $i$ & $>$19.6 & - \\
60055.3 & $g$ & 19.80 & 0.29\\
60057.3 & $g$ & 19.24 & 0.10\\
60058.3 & $i$ & $>$19.4 & - \\
60059.3 & $r$ & 18.86 & 0.11\\
60059.3 & $g$ & 18.75 & 0.09\\
60061.3 & $r$ & 18.50 & 0.10\\
60061.3 & $g$ & 18.28 & 0.10\\
60061.4 & $i$ & 18.62 & 0.09\\
60063.2 & $g$ & 18.00 & 0.07\\
60065.3 & $g$ & 17.64 & 0.08\\
60073.3 & $g$ & 16.71 & 0.04\\
60073.3 & $r$ & 16.99 & 0.06\\
60077.2 & $r$ & 16.75 & 0.04\\
60077.4 & $g$ & 16.59 & 0.06\\
60079.3 & $g$ & 16.41 & 0.04\\
60079.3 & $r$ & 16.67 & 0.04\\
60081.3 & $r$ & 16.60 & 0.04\\
60081.3 & $g$ & 16.35 & 0.04\\
60083.2 & $g$ & 16.32 & 0.04\\
60083.2 & $r$ & 16.56 & 0.04\\
60085.2 & $g$ & 16.32 & 0.03\\
60085.3 & $r$ & 16.52 & 0.04\\
60085.3 & $i$ & 16.75 & 0.03\\
60087.3 & $g$ & 16.33 & 0.05\\
60088.2 & $i$ & 16.73 & 0.03\\
60089.3 & $g$ & 16.35 & 0.04\\
60089.3 & $r$ & 16.53 & 0.04\\
60097.2 & $r$ & 16.67 & 0.04\\
60099.2 & $g$ & 16.59 & 0.04\\
60099.2 & $r$ & 16.65 & 0.05\\
60102.2 & $r$ & 16.73 & 0.04\\
60102.3 & $g$ & 16.76 & 0.06\\
60105.2 & $g$ & 16.78 & 0.04\\
60105.2 & $r$ & 16.81 & 0.04\\
60109.2 & $r$ & 16.88 & 0.04\\
60109.2 & $g$ & 16.92 & 0.04\\
60110.2 & $i$ & 16.99 & 0.02\\
60111.3 & $g$ & 16.97 & 0.05\\
60113.2 & $i$ & 16.99 & 0.02\\
60113.2 & $r$ & 16.97 & 0.05\\
60115.2 & $g$ & 17.11 & 0.06\\
60117.2 & $i$ & 17.03 & 0.02\\
60118.2 & $r$ & 17.01 & 0.05\\
60118.2 & $g$ & 17.21 & 0.06\\
60120.2 & $r$ & 17.10 & 0.05\\
60120.3 & $g$ & 17.32 & 0.06\\
60125.2 & $r$ & 17.17 & 0.06\\
60127.2 & $g$ & 17.51 & 0.08\\
\hline
\end{tabular}
\tablefoot{Host-galaxy subtraction has been performed at all epochs.}
\end{table}

\begin{table}[h!]
\caption{\textit{Swift}/UVOT photometry of SN~2023gpw.}
\label{tab:uvot}
\begin{tabular}{cccc}
\hline
MJD & Filter & Magnitude & Error \\
 & & (mag, AB) & (mag, AB) \\
\hline
60075.8	& UVW2 & 16.37 & 0.03 \\
60075.8 & UVM2 & 16.29 & 0.03 \\
60075.8 & UVW1 & 16.34 & 0.03 \\
60082.9 & UVW2 & 16.55 & 0.02 \\
60082.9 & UVM2 & 16.38 & 0.03 \\
60089.6 & UVW2 & 17.04 & 0.03 \\
60089.6 & UVM2 & 16.75 & 0.04 \\
60089.6 & UVW1 & 16.56 & 0.03 \\
60104.6 & UVW2 & 18.46 & 0.08 \\
60104.6 & UVM2 & 18.21 & 0.09 \\
60104.6 & UVW1 & 17.70 & 0.08 \\
60108.6 & UVW2 & 18.99 & 0.09 \\
60108.6 & UVM2 & 18.49 & 0.09 \\
60108.6 & UVW1 & 18.15 & 0.08 \\
60118.3 & UVW2 & 20.08 & 0.20 \\
60118.3 & UVM2 & 19.37 & 0.16 \\
60118.3 & UVW1 & 18.98 & 0.13 \\
60129.4 & UVW2 & 21.40 & 0.29 \\
60129.4 & UVM2 & 20.88 & 0.28 \\
60129.4 & UVW1 & 20.11 & 0.16 \\
60135.8 & UVW2 & 21.10 & 0.24 \\
60135.8 & UVM2 & 20.67 & 0.22 \\
60135.8 & UVW1 & 20.32 & 0.18 \\
60141.3 & UVW2 & 21.99 & 0.36 \\
60141.3 & UVM2 & 21.41 & 0.32 \\
60141.3 & UVW1 & 21.57 & 0.39 \\
\hline
\end{tabular}
\tablefoot{Host flux subtraction has been performed at all epochs.}
\end{table}

\begin{table*}[h!]
\centering
\caption{Log of our spectroscopic observations of SN~2023gpw. }
\label{tab:specs}
\begin{tabular}{cccccccc}
\hline
MJD & Epoch & Instrument & Slit & Grism/grating & $R$ & Exposure time & Airmass \\
 & (d) &  &  &  &  & (s) & \\
\hline
60071.2 & $-12.7$ & P60/SEDM & IFU & - & 100 & 2250 & 1.3 \\
60073.2 & $-10.9$ & P60/SEDM & IFU & - & 100 & 2250 & 1.4 \\
60075.3 & $-8.9$ & P60/SEDM & IFU & - & 100 & 2250 & 1.7 \\
60079.4 & $-5.2$ & P60/SEDM & IFU & - & 100 & 2250 & 2.2 \\
60081.3 & $-3.4$ & P60/SEDM & IFU & - & 100 & 2250 & 1.9 \\
60084.3 & $-0.7$ & P60/SEDM & IFU & - & 100 & 2250 & 1.3 \\
60098.3 & 12.3 & P60/SEDM & IFU & - & 100 & 1800 & 2.1 \\
60119.2 & 31.6 & P60/SEDM & IFU & - & 100 & 1800 & 1.7 \\
60120.2 & 32.5 & P60/SEDM & IFU & - & 100 & 1800 & 1.9 \\
60121.2 & 33.4 & P60/SEDM & IFU & - & 100 & 1800 & 1.5 \\
\hline
60071.1 & $-12.8$ & NOT/ALFOSC & $1\farcs0$ & \#4 & 360 & 600 & 1.7 \\
60077.0 & $-7.4$ & NOT/ALFOSC & $1\farcs0$ & \#4 & 360 & 900 & 1.3 \\
60084.9 & $-0.1$ & NOT/ALFOSC & $1\farcs0$ & \#4 & 360 & 900 & 1.2 \\
\hline
60097.0 & 11.1 & INT/IDS & $1\farcs5$ & R300V & 980 & 1800 & 1.6 \\
60097.9 & 11.9 & INT/IDS & $1\farcs5$ & R300V & 980 & 2700 & 1.2 \\
\hline
60116.2 & 28.8 & P200/DBSP & $1\farcs0$ & 600/4000 & 1376 & 300 & 1.5 \\
60116.2 & 28.8 & P200/DBSP & $1\farcs0$ & 316/7500 & 1368 & 300 & 1.5 \\
60136.2 & 47.3 & P200/DBSP & $1\farcs5$ & 600/4000 & 917 & 600 & 1.8 \\
60136.2 & 47.3 & P200/DBSP & $1\farcs5$ & 316/7500 & 912 & 600 & 1.8 \\
\hline
60386.2 & 278.2 & GTC/OSIRIS+ & $1\farcs0$ & R1000B & 610 & 1800 & 1.3 \\
60386.2 & 278.2 & GTC/OSIRIS+ & $1\farcs0$ & R1000R & 670 & 1800 & 1.3 \\
\hline
\end{tabular}
\tablefoot{The resolving power ($R$) is reported at the blaze wavelength of each grism/grating. The epochs are reported as rest-frame days since the $g$-band peak on MJD\,=\,60085.0.}
\end{table*}

\begin{table}
\centering
\caption{Photometry of the host galaxy of SN~2023gpw based on publicly available pre-explosion data and our UVOT measurements.} 
\label{tab:host}
\begin{tabular}{lccc}
\hline
Source & Filter & Magnitude & Error \\
\hline
\textit{Swift}/UVOT & UVW2   &$ 22.55$ & $0.52 $\\
\textit{Swift}/UVOT & UVM2   &$ 23.30$ & $1.02 $\\
\textit{Swift}/UVOT & UVW1   &$ 21.47$ & $0.31 $\\
LS & $g$        &$ 20.63$ & $0.05 $\\
LS & $r$        &$ 20.16 $ & $ 0.04 $\\
LS & $i$        &$ 20.10 $ & $ 0.03 $\\
LS & $z$        &$ 19.95 $ & $ 0.11 $\\
PS1 & $g$       &$ 20.64 $ & $ 0.12 $\\
PS1 & $i$       &$ 20.20 $ & $ 0.10 $\\
PS1 & $z$       &$ 19.90 $ & $ 0.12 $\\
\textit{WISE} & W1     &$ 20.57 $ & $ 0.28 $\\
\textit{WISE} & W2     &$ 21.85 $ & $ 0.80 $\\
\hline
\end{tabular}
\end{table}

\FloatBarrier

\section{MOSFiT corner plots}
\label{sec:corners}

\begin{figure}[h!]
\centering
\includegraphics[width=1.8\linewidth]{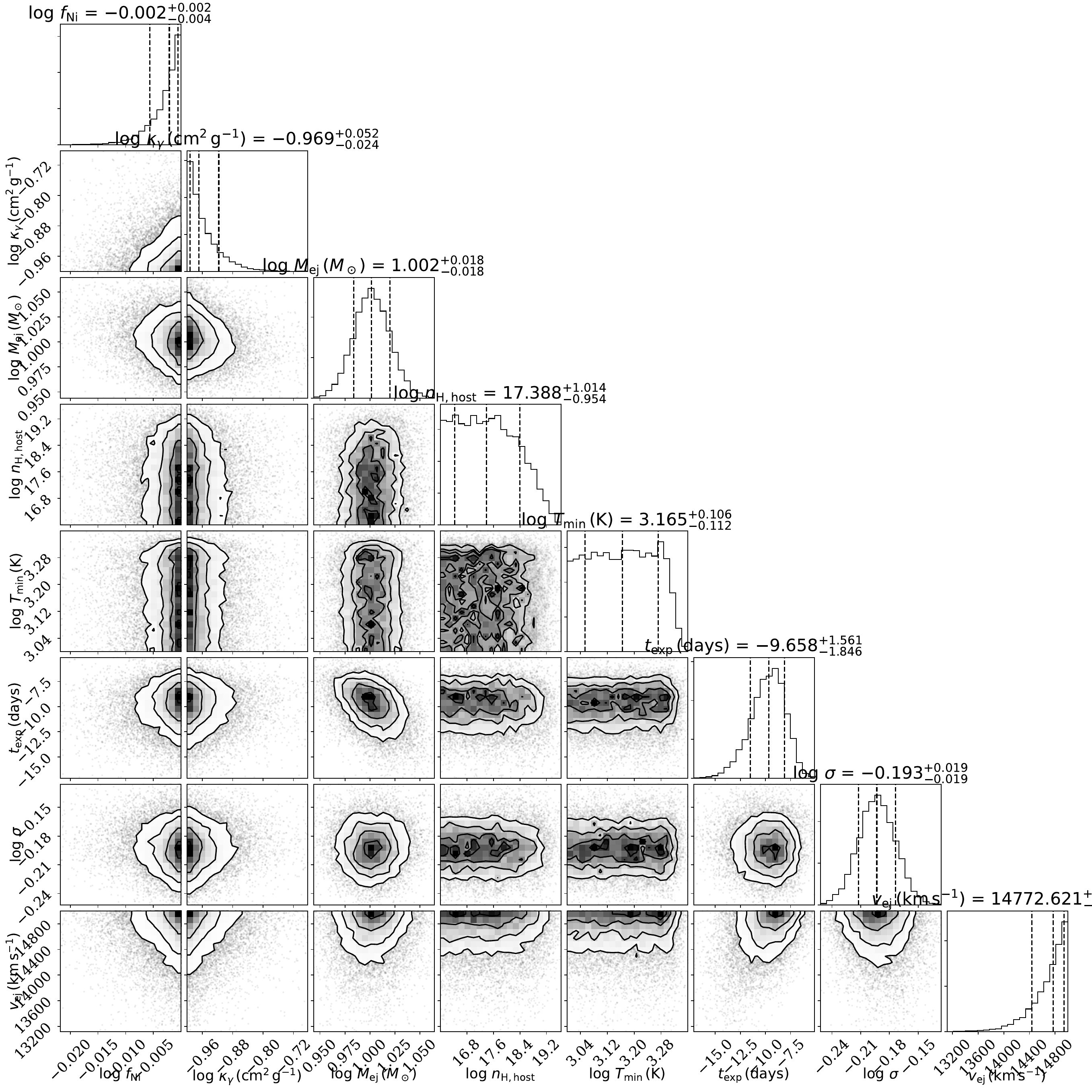}
\caption{\texttt{MOSFiT} corner plot for the $^{56}$Ni decay fit.}
\label{fig:cornerni56}
\end{figure}

\onecolumn

\begin{figure}
\centering
\includegraphics[width=0.9\linewidth]{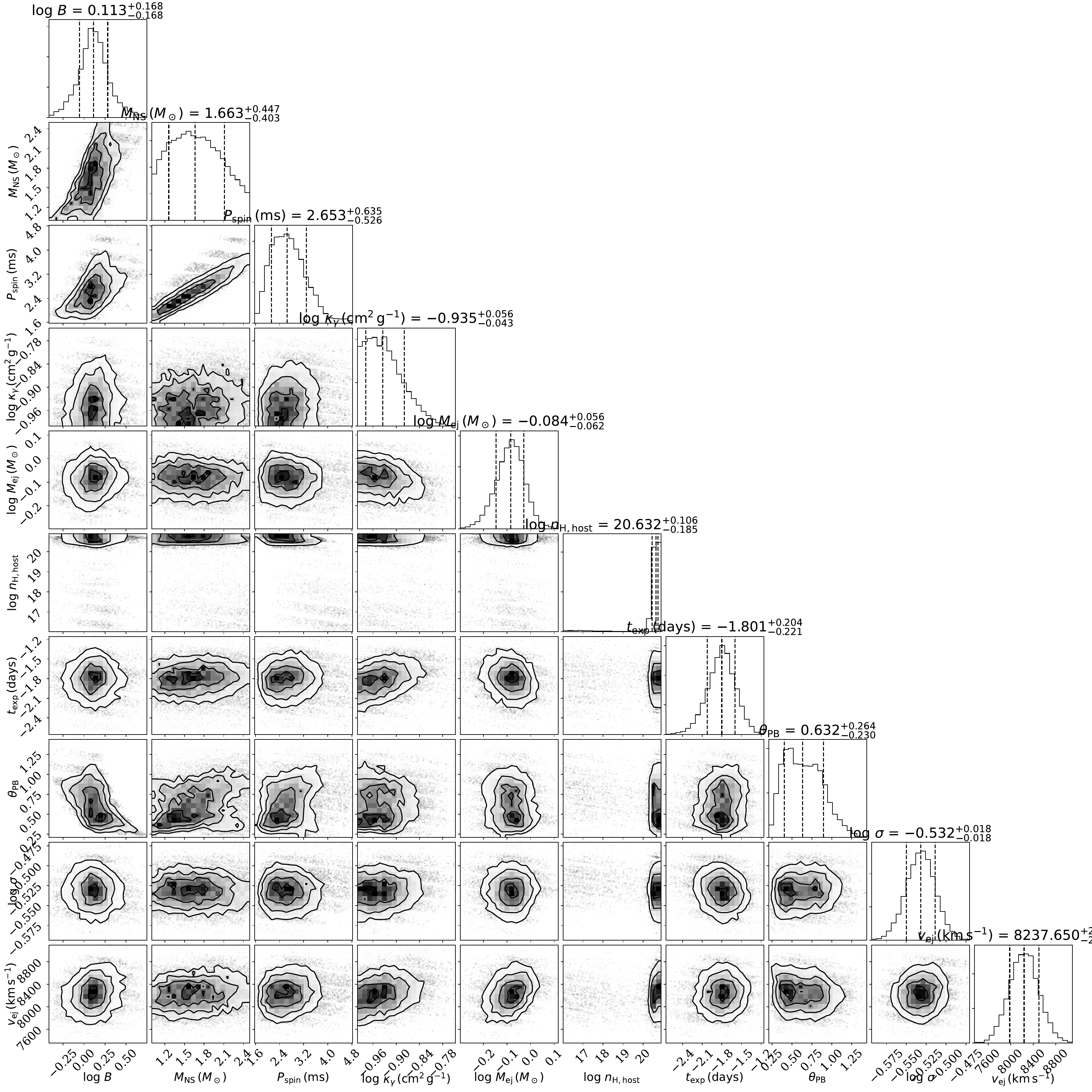}
\caption{\texttt{MOSFiT} corner plot for the magnetar spindown fit.}
\label{fig:cornermag}
\end{figure}

\begin{figure}
\centering
\includegraphics[width=0.9\linewidth]{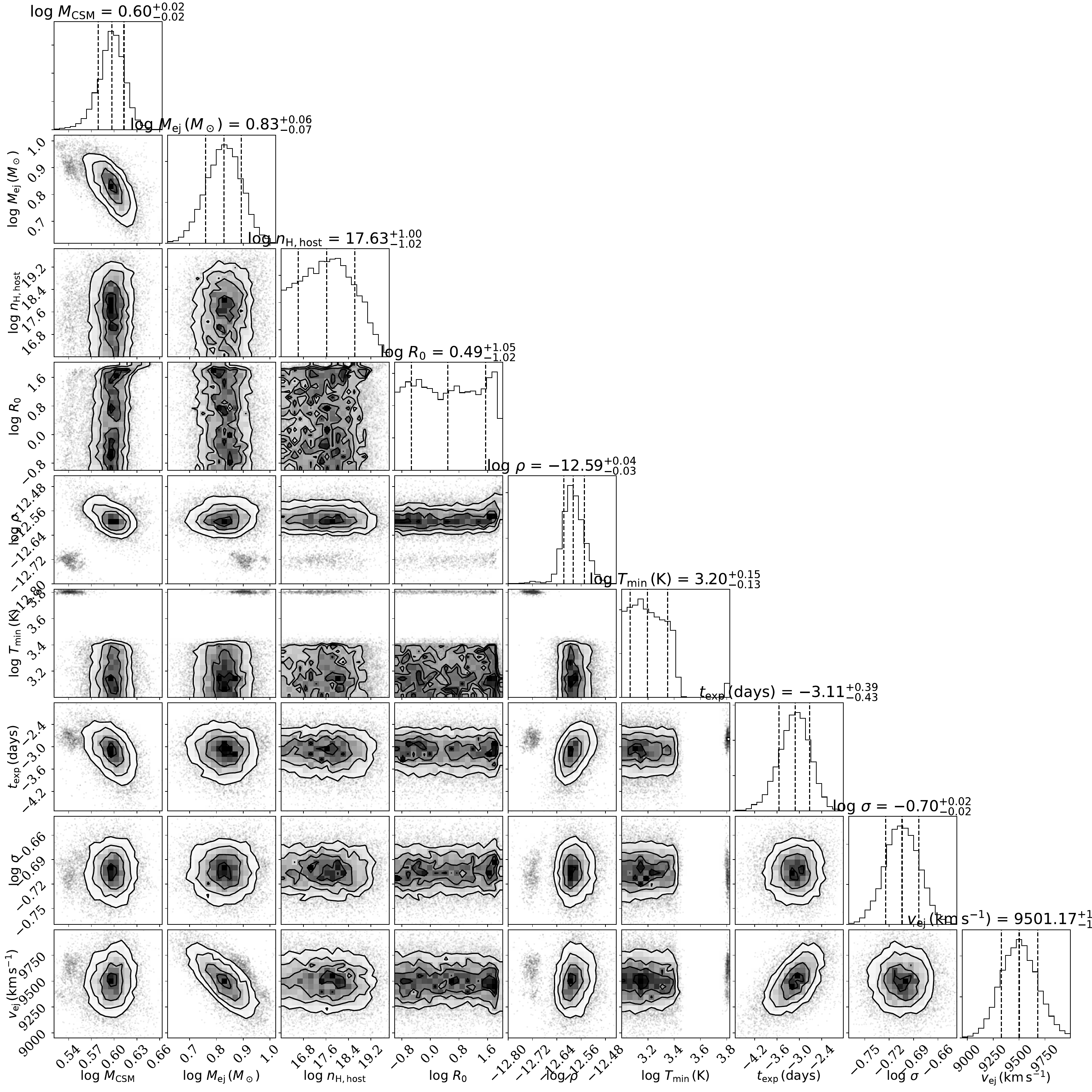}
\caption{\texttt{MOSFiT} corner plot for the CSI fit when $s=0$.}
\label{fig:cornercms0}
\end{figure}

\begin{figure}
\centering
\includegraphics[width=0.9\linewidth]{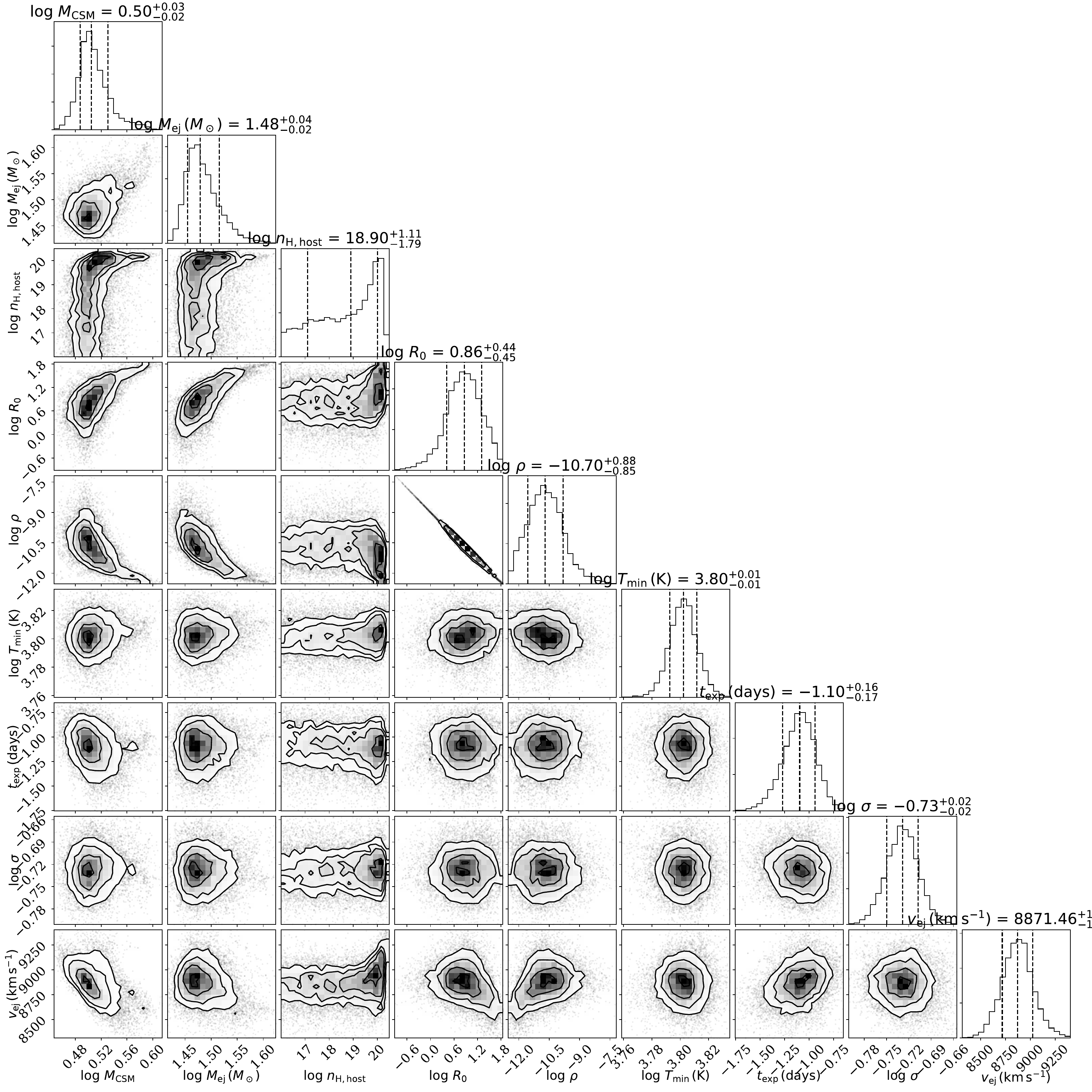}
\caption{\texttt{MOSFiT} corner plot for the CSI fit when $s=2$.}
\label{fig:cornercsms2}
\end{figure}

\begin{figure}
\centering
\includegraphics[width=0.9\linewidth]{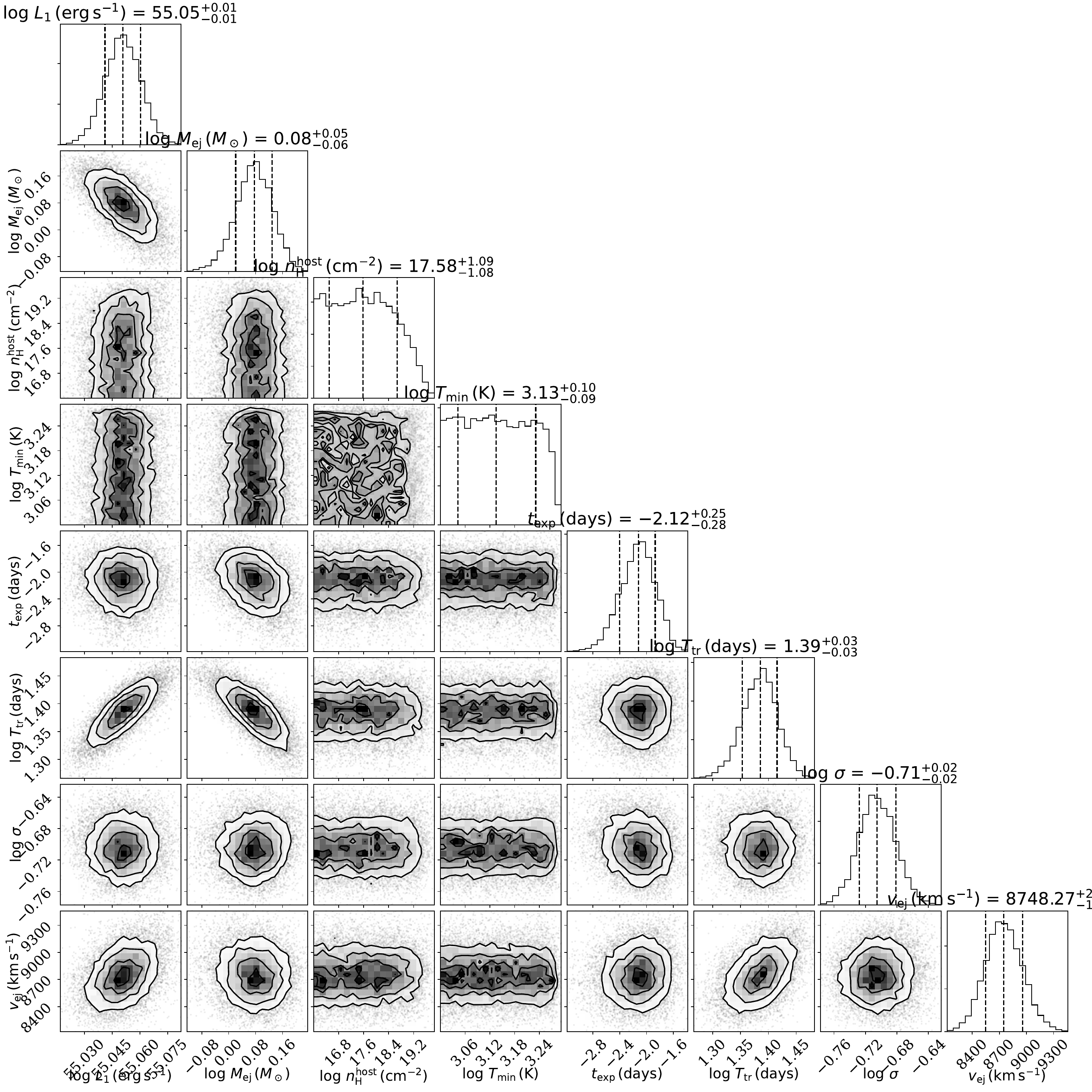}
\caption{\texttt{MOSFiT} corner plot for the fallback fit.}
\label{fig:cornerFB}
\end{figure}
    
\end{appendix}

\end{document}